\documentclass[aps,prb,twocolumn,showpacs,groupedaddress]{revtex4}

\usepackage{graphicx}
\usepackage{amssymb}

\begin{document}

% number in the upper righthand corner of the title page in preprint 
% mode.
%\preprint{}

\title{Nonlinear magnetic susceptibility and aging phenomena in reentrant 
ferromagnet: Cu$_{0.2}$Co$_{0.8}$Cl$_{2}$-FeCl$_{3}$ graphite 
bi-intercalation compound}


\author{Masatsugu Suzuki }
\email[]{suzuki@binghamton.edu}
%\thanks{}
\affiliation{Department of Physics, State University of New York at 
Binghamton, Binghamton, New York 13902-6016}

\author{Itsuko S. Suzuki }
\email[]{itsuko@binghamton.edu}
%\thanks{}
\affiliation{Department of Physics, State University of New York at 
Binghamton, Binghamton, New York 13902-6016}


\date{\today}

\begin{abstract}
Linear and nonlinear dynamic properties of a reentrant ferromagnet
Cu$_{0.2}$Co$_{0.8}$Cl$_{2}$-FeCl$_{3}$ graphite bi-intercalation compound
are studied using AC and DC magnetic susceptibility.  This compound
undergoes successive phase transitions at the transition temperatures
$T_{h}$ (= 16 K), $T_{c}$ (= 9.7 K), and $T_{RSG}$ (= 3.5 K).  The static
and dynamic behaviors of the reentrant spin glass phase below $T_{RSG}$ are
characterized by those of normal spin glass phase with critical exponent
$\beta$ = 0.57 $\pm$ 0.10, a dynamic critical exponent $x$ = 8.5 $\pm$ 1.8,
and an exponent $p$ (= 1.55 $\pm$ 0.13) for the de Almeida -Thouless line. 
A prominent nonlinear susceptibility is observed between $T_{RSG}$ and
$T_{c}$ and around $T_{h}$, suggesting a chaotic nature of the
ferromagnetic phase ($T_{RSG} \leq T \leq T_{c}$) and the helical spin
ordered phase ($T_{c} \leq T \leq T_{h}$).  The aging phenomena are
observed both in the RSG and FM phases, with the same qualitative features
as in normal spin glasses.  The aging of zero-field cooled magnetization
indicates a drastic change of relaxation mechanism below and above
$T_{RSG}$.  The time dependence of the absorption $\chi^{\prime \prime}$ is
described by a power law form ($\approx t^{-b^{\prime \prime}}$) in the
ferromagnetic phase, where $b^{\prime \prime} \approx 0.074 \pm 0.016$ at
$f$ = 0.05 Hz and $T$ = 7 K. No $\omega t$-scaling law for $\chi^{\prime
\prime}$ [$\approx (\omega t)^{-b^{\prime \prime}}$] is observed.
\end{abstract}

%\keywords{}

\maketitle

%\section{}
%\subsection{}
%\subsubsection{}

% If in two-column mode, this environment will change to single-column
% format so that long equations can be displayed. Use
%\begin{widetext}
%\end{widetext}

\section{\label{intro}Introduction}
The nature of a reentrant spin glass (RSG) phase and a ferromagnetic (FM)
phase in reentrant ferromagnets has been a topic of much controversy. 
\cite{Motoya1983,SuzukiJ1990,Geohegan1981,Jonason1996a,Jonason1996b,
Aeppli1983,Sato2001,Ogawa2001,Dormann1985,Alba1986,Pouget1995,Dupuis2002,
Maletta1982,Aeppli1987,Mathieu2001} 
Spin frustration effects occur as a result of a competition between
ferromagnetic interactions as a majority and antiferromagnetic interactions
as a minority.  As the temperature is lowered, the reentrant ferromagnet
exhibits first a transition from a paramagnetic (PM) phase to a FM phase
with decreasing temperature and then a second transition from the FM phase
to a RSG phase.  Such a reentry behavior of the reentrant ferromagnets has
been extensively studied experimentally in the last two decades.
\cite{Motoya1983,SuzukiJ1990,Geohegan1981,Jonason1996a,Jonason1996b,
Aeppli1983,Sato2001,Ogawa2001,Dormann1985,Alba1986,Pouget1995,Dupuis2002,
Maletta1982,Aeppli1987,Mathieu2001} 
There are four types of reentrant ferromagnets: (i) metallic spin glasses
such as Fe$_{0.7}$Al$_{0.3}$,\cite{Motoya1983,SuzukiJ1990}
(Fe$_{0.20}$Ni$_{0.80}$)$_{75}$P$_{16}$B$_{6}$Al$_{3}$,
\cite{Geohegan1981,Jonason1996a,Jonason1996b}
(Fe$_{1-x}$Mn$_{x}$)$_{75}$P$_{16}$B$_{6}$Al$_{3}$ ($0.2 \leq x \leq
0.32$),\cite{Aeppli1983} and N$_{77}$Mn$_{22}$\cite{Sato2001,Ogawa2001}
having Ruderman-Kittel-Kasuya-Yosida (RKKY)-type interactions between
distant spins, (ii) insulating spin glasses such as
CdCr$_{2x}$In$_{2(1-x)}$S$_{4}$ ($0.90 \leq x <
1$),\cite{Dormann1985,Alba1986,Pouget1995,Dupuis2002} (iii) dilute magnetic
semiconductors such as Eu$_{x}$Sr$_{1-x}$S ($0.52 \leq x \leq
0.60$),\cite{Maletta1982,Aeppli1987} and (iv) colossal magnetoresistance
materials such as Y$_{0.7}$Ca$_{0.3}$MnO$_{3}$.\cite{Mathieu2001} In the
case of (ii) and (iii), ferromagnetic nearest neighbor interactions compete
with antiferromagnetic next nearest neighbor interactions.

The nature of the RSG and FM phases in reentrant ferromagnets is
basically understood in terms of a mean-field picture.  The phase diagram
($T/J$ vs $J_{0}/J$) of the Sherrington-Kirkpatrick (SK) model with Ising
spins,\cite{Sherrington1975} consists of the PM phase, FM phase, and the
spin glass (SG) phase, where an infinite-ranged Gaussian distribution of
exchange interactions with variance $J$ and mean $J_{0}$ is
assumed.\cite{Edwards1975} For $J_{0}/J \leq 1$, as $T$ is lowered, the
system undergoes a transition from the PM phase to the SG phase.  For
$J_{0}/J > 1$, as $T$ is lowered, the system undergoes a PM-FM transition
followed by a FM-SG transition.  This SG phase for $J_{0}/J > 1$ is called
a RSG phase.  However, the nature of the RSG phase is essentially the same as
that of the normal SG phase for $J_{0}/J \leq 1$.  When the Parisi's
solution for the SK model\cite{Parisi1979,Toulouse1980} is discovered, the
vertical phase boundary at $J_{0}/J$ = 1 is added to the SK phase diagram. 
For $J_{0}/J > 1$, consequently the whole RSG phase and a part of the FM
phase in the SK model are newly replaced by a RSG phase with replica
symmetry breaking (RSB).  This RSG phase is very different from the normal
SG phase for $J_{0}/J \leq 1$.  It is a mixed phase of the SG phase and the
FM phase.

In the mean-field picture, a true reentrance from the FM phase to the
normal SG phase is not predicted.  There is a normal FM long-range order in
the FM phase.  This picture, which assumes infinite-range interactions, is
not always appropriate for real reentrant magnets where the short-range
interactions are large and random in sign and the spin symmetry is rather
Heisenberg-like than Ising-like.  Neutron scattering studies on
(Fe$_{1-x}$Mn$_{x}$)$_{75}$P$_{16}$B$_{6}$Al$_{3}$\cite{Aeppli1983} have
questioned the existence of a true long-range order even in the FM phase. 
Aeppli et al.\cite{Aeppli1983} have proposed a phenomenological
random-field picture to explain their result.  In this picture, the system
in the FM phase consists of regions which would order
ferromgnetically and other regions forming PM clusters.  The frustrated
spins in the PM clusters can generate random molecular fields, which act on
the unfrustrated spins in the infinite FM network.  In the FM phase well
above $T_{RSG}$, the fluctuations of the spins in the PM clusters are so
rapid that the FM network is less influenced by them and their effect is
only to reduce the net FM moment.  On decreasing the temperature toward
$T_{RSG}$, the thermal fluctuations of the spins in the PM clusters become
slower.  The coupling between the PM clusters and the FM network becomes
important and the molecular field from the slow PM spins acts as a random
magnetic field.  This causes breakups of the FM network into finite-sized
clusters.  Below $T_{RSG}$, the ferromagnetism completely disappears,
leading to a SG phase.  This picture is very different from the mean-field
picture.  The RSG phase is not a mixed phase but a normal SG phase.

Jonason et al.\cite{Jonason1996a,Jonason1996b} have shown from a dynamic
scaling analysis of low-field dynamic susceptibility of
(Fe$_{0.20}$Ni$_{0.80}$)$_{75}$P$_{16}$B$_{6}$Al$_{3}$ that there is a
spin-glass relaxation time which diverges at a finite temperature with a
dynamic critical exponent similar to that observed for an ordinary PM-SG
transition.  This suggests that the RSG phase is a normal SG phase.  The FM
phase just above $T_{RSG}$ shows a dynamic behavior characterized by an
aging effect and chaotic nature similar to that of SG phase.  Dupuis et
al.\cite{Dupuis2002} have shown that the aging behavior of the low
frequency AC susceptibility is observed both in the FM phase and RSG phase
of CdCr$_{2x}$In$_{2(1-x)}$S$_{4}$ with $x$ = 0.90, 0.95, and 1.00, with
the same qualitative features as in normal spin glasses.  It seems that
these results are explained in terms of the random-field picture.

In this paper we report our experimental study on the magnetic properties
of the RSG phase and the FM phase of a reentrant ferromagnet
Cu$_{0.2}$Co$_{0.8}$Cl$_{2}$-FeCl$_{3}$ graphite bi-intercalation compound
(GBIC).  This compound
undergoes three phase transitions at $T_{h}$ (= 16 K), $T_{c}$ (= 9.7 K),
and $T_{RSG}$ (= 3.5 K).  There are the helical spin ordered phase between
$T_{h}$ and $T_{c}$, the FM phase between $T_{c}$ and $T_{RSG}$, and the
RSG phase below
$T_{RSG}$.\cite{Suzuki1997,Suzuki2000,Suzuki1999,Suzuki2003} The static and
dynamic natures of the FM and RSG phases are examined from the linear and
nonlinear AC magnetic susceptibility, the magnetizations in the zero-field
cooled (ZFC), field-cooled (FC), isothermal remnant (IR), and thermoremnant
(TR) states.  The time dependence of the AC susceptibility and the ZFC
magnetization, showing the aging phenomena, is also studied.

Cu$_{0.2}$Co$_{0.8}$Cl$_{2}$-FeCl$_{3}$ GBIC has a unique layered structure
where the Cu$_{0.2}$Co$_{0.8}$Cl$_{2}$ intercalate layer (= $I_{1}$) and
FeCl$_{3}$ intercalate layers (= $I_{2}$) alternate with a single graphite
layer ($G$), forming a stacking sequence
(-$G$-$I_{1}$-$G$-$I_{2}$-$G$-$I_{1}$-$G$-$I_{2}$-$G$-$\cdots$) along the
$c$ axis.  In the Cu$_{0.2}$Co$_{0.8}$Cl$_{2}$ intercalate layer two kinds
of magnetic ions (Cu$^{2+}$ and Co$^{2+}$) are randomly distributed on the
triangular lattice.  The spin order in the Cu$_{0.2}$Co$_{0.8}$Cl$_{2}$
layers is coupled with that in the FeCl$_{3}$ intercalate layer (= $I_{2}$)
through an interplanar exchange interaction, leading to the spin
frustration effect.

\section{\label{exp}EXPERIMENTAL PROCEDURE}
A sample of stage-2 Cu$_{0.2}$Co$_{0.8}$Cl$_{2}$ graphite intercalation
compound (GIC) as a starting material was prepared from single crystal kish
graphite (SCKG) by vapor reaction of anhydrated
Cu$_{0.2}$Co$_{0.8}$Cl$_{2}$ in a chlorine atmosphere with a gas pressure
of $\approx$ 740 Torr.\cite{Suzuki1997,Suzuki2000,Suzuki1999,Suzuki2003}
The reaction was continued at 500 $^\circ$C for three weeks.  The sample of
Cu$_{0.2}$Co$_{0.8}$Cl$_{2}$-FeCl$_{3}$ GBIC was prepared by a sequential
intercalation method: the intercalant FeCl$_{3}$ was intercalated into
empty graphite galleries of stage-2 Cu$_{0.2}$Co$_{0.8}$Cl$_{2}$ GIC. A
mixture of well-defined stage-2 Cu$_{0.2}$Co$_{0.8}$Cl$_{2}$ GIC based on
SCKG and single-crystal FeCl$_{3}$ was sealed in vacuum inside Pyrex glass
tubing, and was kept at 330 $^\circ$C for two weeks.  The stoichiometry of
the sample is represented by
C$_{m}$(Cu$_{0.2}$Co$_{0.8}$Cl$_{2}$)$_{1-x}$(FeCl$_{3}$)$_{x}$.  The
concentration of C and Fe ($m$ and $x$) was determined from weight uptake
measurement and electron microprobe measurement [using a scanning electron
microscope (Model Hitachi S-450)]: $m = 7.54 \pm 0.05$ and $x = 0.56 \pm
0.02$.  The (00$L$) x-ray diffraction measurements of stage-2
Cu$_{0.2}$Co$_{0.8}$Cl$_{2}$ GIC and
Cu$_{0.2}$Co$_{0.8}$Cl$_{2}$-FeCl$_{3}$ GBIC were made at 300 K by using a
Huber double circle diffractometer with a MoK$\alpha$ x-ray radiation
source (1.5 kW).  The $c$-axis repeat distance of stage-2
Cu$_{0.2}$Co$_{0.8}$Cl$_{2}$ GIC and
Cu$_{0.2}$Co$_{0.8}$Cl$_{2}$-FeCl$_{3}$ GBIC was determined as 12.81 $\pm$
0.04 and 18.79 $\pm$ 0.05 $\AA$, respectively.

The DC magnetization and AC susceptibility were measured using a SQUID
magnetometer (Quantum Design, MPMS XL-5) with an ultra low-field capability
option.  First a remnant magnetic field was reduced to zero field (exactly
less than 3 mOe) at 298 K for both DC magnetization and AC susceptibility
measurements.  Then the sample was cooled from 298 to 1.9 K in a zero
field.  (i) \textit{Measurements of the zero field cooled susceptibility
($\chi_{ZFC}$) and the field cooled susceptibility ($\chi_{FC}$)}.  After
an external magnetic field $H$ (0 $\leq H \leq$ 5 kOe) was applied along
the $c$ plane (basal plane of graphene layer) at 1.9 K, $\chi_{ZFC}$ was
measured with increasing temperature ($T$) from 1.9 to 20 K. After
annealing of sample for 10 minutes at 100 K in the presence of $H$,
$\chi_{FC}$ was measured with decreasing $T$ from 20 to 1.9 K. (ii)
\textit{AC susceptibility measurement}.  The frequency ($f$), magnetic
field, and temperature dependence of the dispersion
($\Theta_{1}^{\prime}/h$) and absorption ($\Theta_{1}^{\prime \prime}/h$)
was measured between 1.9 to 20 K, where the frequency of the AC field is
$f$ = 0.01 - 1000 Hz and the amplitude $h$ is typically $h$ = 1 mOe - 4.2
Oe.

\section{\label{result}RESULT}
\subsection{\label{resultA}Nonlinear AC susceptibility: 
$\Theta_{1}^{\prime}/h$ and $\Theta_{1}^{\prime \prime}/h$}

\begin{figure}
\includegraphics[width=7.5cm]{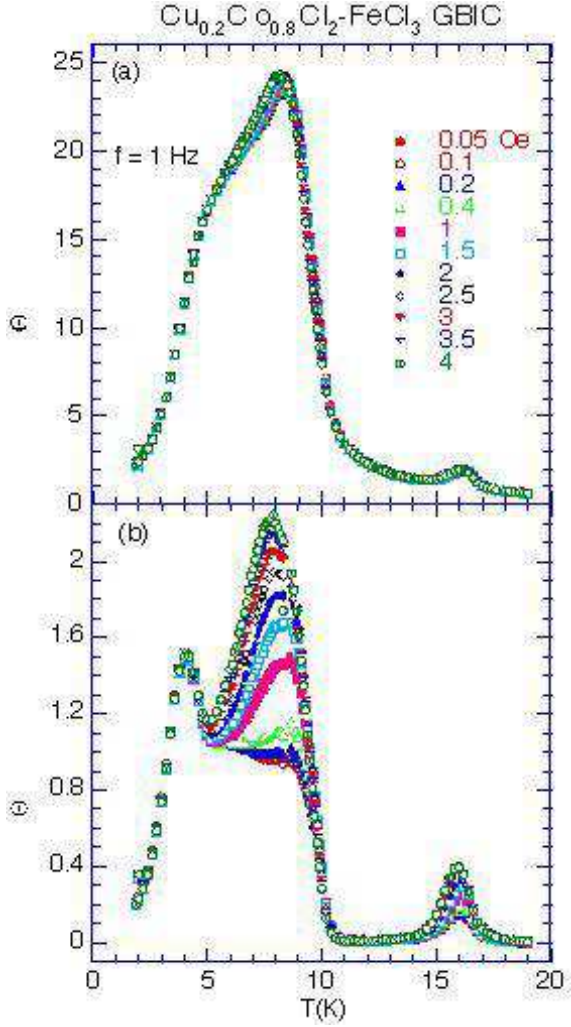}
\caption{\label{fig:one}$T$ dependence of (a) $\Theta_{1}^{\prime}/h$ and
(b) $\Theta_{1}^{\prime \prime}/h$ at various $h$ for
Cu$_{0.2}$Co$_{0.8}$Cl$_{2}$-FeCl$_{3}$ GBIC. $h \perp c$.  $f$ = 1 Hz. 
$H$ = 0.}
\end{figure}

We have measured the dispersion $\Theta_{1}^{\prime}/h$ and absorption
$\Theta_{1}^{\prime \prime}/h$ at fixed $T$ as a function of $h$, where 1
mOe $\leq h \leq$ 4.2 Oe and $f$ = 1 Hz.  Both $\Theta_{1}^{\prime}/h$ and
$\Theta_{1}^{\prime \prime}/h$ are strongly dependent on $h$, suggesting
the existence of nonlinear magnetic susceptibility.  We determine the $T$
dependence of nonlinear AC magnetic susceptibilities ($\chi_{3}^{\prime}$,
$\chi_{5}^{\prime}$, $\chi_{3}^{\prime \prime}$, $\chi_{5}^{\prime
\prime}$) from the least squares fits of the data to the power law
forms:\cite{Suzuki2002}
\begin{equation} 
\Theta_{1}^{\prime}/h = \chi_{1}^{\prime} + 3\chi_{3}^{\prime}h^{2}/4 + 
10\chi_{5}^{\prime}h^{4}/16 + \cdots,
\label{eq:one} 
\end{equation} 
and
\begin{equation} 
\Theta_{1}^{\prime \prime}/h = \chi_{1}^{\prime \prime} + 
3\chi_{3}^{\prime \prime}h^{2}/4 + 
10\chi_{5}^{\prime \prime}h^{4}/16 + \cdots.
\label{eq:two} 
\end{equation} 
Note that for convenience we use the linear AC susceptibility
$\chi^{\prime}$ and $\chi^{\prime \prime}$ instead of $\chi_{1}^{\prime}$
and $\chi_{1}^{\prime \prime}$.  In Figs.~\ref{fig:one}(a) and (b) we show
the $T$ dependence of the dispersion $\Theta_{1}^{\prime}/h$ and the
absorption $\Theta_{1}^{\prime \prime}/h$ at $f$ = 1 Hz.  The different
curves correspond to different amplitudes of the AC field, where $h$ is
varied between 1 mOe and 4.2 Oe.  The susceptibility
$\Theta_{1}^{\prime}/h$ and $\Theta_{1}^{\prime \prime}/h$ are independent
of $h$ for $h < h_{0}$ within experimental error, where $h_{0} \approx$ 0.1
Oe, implying that $\Theta_{1}^{\prime}/h$ and $\Theta_{1}^{\prime
\prime}/h$ coincide with the linear susceptibilities $\chi^{\prime}$ and
$\chi^{\prime \prime}$, respectively.  Note that $\Theta_{1}^{\prime
\prime}/h$ shows a sharp peak at 4 K, which is almost independent of $h$. 
This sharp peak is associated with the transition between the FM and RSG
phases: $T_{RSG}$ = 3.5 K.

\begin{figure}
\includegraphics[width=7.5cm]{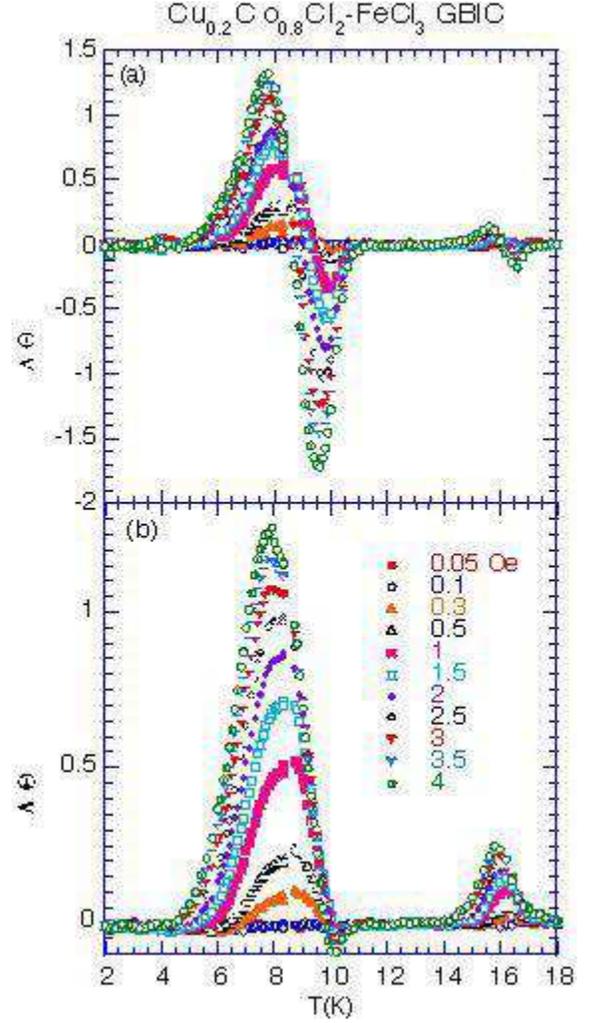}
\caption{\label{fig:two}$T$ dependence of (a)
$\Delta(\Theta_{1}^{\prime}/h)$ and (b)
$\Delta(\Theta_{1}^{\prime\prime}/h)$ at various $h$.  $f$ = 1 Hz.  $H$ =
0.  $\Delta(\Theta_{1}^{\prime}/h)$ is defined by the difference between
$\Theta_{1}^{\prime}/h$ at $h$ ($h>$ 30 mOe) and $\Theta_{1}^{\prime}/h$ at
$h$ = 30 mOe.  $\Delta(\Theta_{1}^{\prime \prime}/h)$ is defined by the
difference between $\Theta_{1}^{\prime \prime}/h$ at $h$ ($h>$ 30 mOe) and
$\Theta_{1}^{\prime \prime}/h$ at $h$ = 30 mOe.}
\end{figure}

In Figs.~\ref{fig:two}(a) and (b) we show the $T$ dependence of
$\Delta(\Theta_{1}^{\prime}/h)$ and $\Delta(\Theta_{1}^{\prime \prime}/h)$
at various $h$, where $\Delta(\Theta_{1}^{\prime}/h)$ and
$\Delta(\Theta_{1}^{\prime \prime}/h)$ are defined as the differences
between $\Theta_{1}^{\prime}/h$ and $\Theta_{1}^{\prime \prime}/h$ at $h$
($h>$ 30 mOe) and those at $h$ = 30 mOe, respectively:
$\Delta(\Theta_{1}^{\prime}/h) = 3\chi_{3}^{\prime}h^{2}/4 +
10\chi_{5}^{\prime}h^{4}/16 +\cdots$ and $\Delta(\Theta_{1}^{\prime
\prime}/h) = 3\chi_{3}^{\prime \prime}h^{2}/4 + 10\chi_{5}^{\prime
\prime}h^{4}/16 +\cdots$.  The differences $\Delta(\Theta_{1}^{\prime}/h)$
and $\Delta(\Theta_{1}^{\prime \prime}/h)$ are strongly dependent on $h$. 
The difference $\Delta(\Theta_{1}^{\prime \prime}/h)$ has two local maxima
at 16.2 K (15.8 K) and 8.6 K (7.8 K), and a local minimum at 10.3 K (10.1 K)
at $h$ = 1 Oe ($h$ = 4 Oe).  In contrast, $\Delta(\Theta_{1}^{\prime}/h)$
has two local maxima at 15.8 K (15.6 K) and 8.3 K (7.8 K), and two local
minima at 9.9 K (9.5 K) and 16.6 K (16.5 K) at $h$ = 1 Oe ($h$ = 4 Oe).  In
summary, the positions of the local maxima and minima shift to low-$T$
side as $h$ increases.

\begin{figure}
\includegraphics[width=7.5cm]{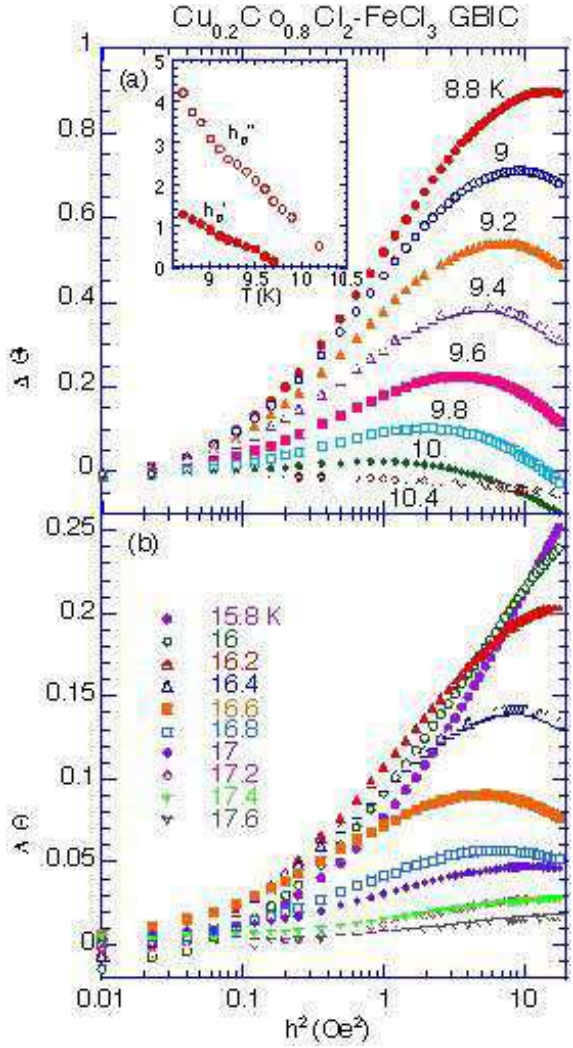}
\caption{\label{fig:three}(a) and (b) $h^{2}$ dependence of
$\Delta(\Theta_{1}^{\prime \prime}/h)$ at various $T$.  $f$ = 1 Hz.  The
inset of (a) shows the $T$ dependence of the peak field $h_{p}^{\prime}$
for $\Delta(\Theta_{1}^{\prime}/h)$ and $h_{p}^{\prime \prime}$ for
$\Delta(\Theta_{1}^{\prime \prime}/h)$.}
\end{figure}

\begin{figure}
\includegraphics[width=7.5cm]{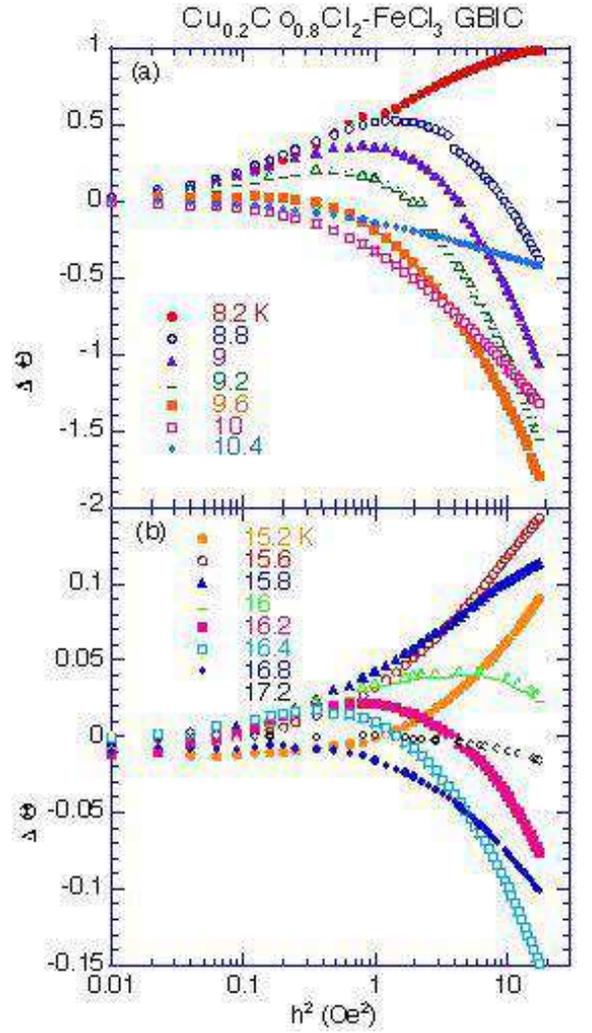}
\caption{\label{fig:four}(a) and (b) $h^{2}$ dependence of
$\Delta(\Theta_{1}^{\prime}/h)$ at various $T$.  $f$ = 1 Hz.}
\end{figure}

In Figs.~\ref{fig:three} and \ref{fig:four} we show the plot of
$\Delta(\Theta_{1}^{\prime \prime}/h)$ and $\Delta(\Theta_{1}^{\prime}/h)$
as a function of $h^{2}$ near $T_{c}$ (= 9.7 K) and $T_{h}$ (= 16 K),
respectively.  Both $\Delta(\Theta_{1}^{\prime \prime}/h)$ and
$\Delta(\Theta_{1}^{\prime}/h)$ are strongly dependent on $h^{2}$ in these
limited temperature regions.  A linear increase of
$\Delta(\Theta_{1}^{\prime}/h)$ and $\Delta(\Theta_{1}/h)$ with $h^{2}$ at
low $h$ indicates a positive sign of $\chi_{3}^{\prime}$ and
$\chi_{3}^{\prime \prime}$.  We find a peak in
$\Delta(\Theta_{1}^{\prime}/h)$ [$\Delta(\Theta_{1}^{\prime \prime}/h)$] at
a peak field $h_{p}^{\prime}$ [$h_{p}^{\prime \prime}$], which shifts to the
low-$h$ side with increasing $T$.  The fields $h_{p}^{\prime}$ and
$h_{p}^{\prime \prime}$ are defined as d($\Theta_{1}^{\prime}/h)/$d$h$ = 0
and d($\Theta_{1}^{\prime \prime}/h)$/d$h$ = 0, respectively: $h_{p}^{\prime}
\approx -3\chi_{3}^{\prime}/(5\chi_{5}^{\prime})$ and $h_{p}^{\prime \prime}
\approx -3\chi_{3}^{\prime \prime}/(5\chi_{5}^{\prime \prime})$.  The
existence of $h_{p}^{\prime}$ and $h_{p}^{\prime \prime}$ indicates a
negative sign of $\chi_{5}^{\prime}$ and $\chi_{5}^{\prime \prime}$.  In
the inset of Fig.~\ref{fig:three}(a), we show the $T$ dependence of
$h_{p}^{\prime}$ and $h_{p}^{\prime \prime}$.  The peak fields
$h_{p}^{\prime}$ and $h_{p}^{\prime \prime}$, which decrease with increasing
$T$, tend to reduce to zero around 9.8 and 10.4 K, respectively.

\begin{figure}
\includegraphics[width=7.5cm]{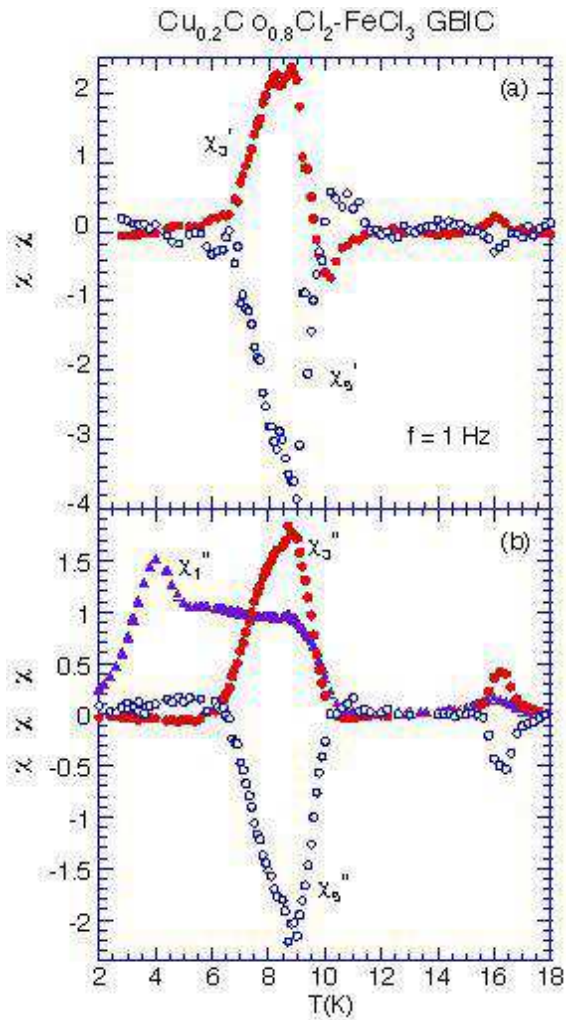}
\caption{\label{fig:five}(a) $T$ dependence of $\chi_{3}^{\prime}$ and
$\chi_{5}^{\prime}$, where the units of $\chi_{2n+1}^{\prime}$ are emu/(av
mol Oe$^{2n+1}$).  $f$ = 1 Hz.  (b) $T$ dependence of
$\chi_{1}^{\prime\prime}$, $\chi_{3}^{\prime\prime}$, and
$\chi_{5}^{\prime\prime}$, where the units of $\chi_{2n+1}^{\prime\prime}$
are emu/(av mol Oe$^{2n+1}$).}
\end{figure}

The least squares fits of the data ($\Theta_{1}^{\prime}/h$ vs $h^{2}$ and
$\Theta_{1}^{\prime\prime}/h$ vs $h^{2}$) at each $T$ for 30 mOe $\leq h
\leq$ 0.7 Oe to Eqs.(\ref{eq:one}) and (\ref{eq:two}) yield the values of
$\chi_{1}^{\prime}$, $\chi_{3}^{\prime}$, $\chi_{5}^{\prime}$,$\cdots$, and
$\chi_{1}^{\prime\prime}$, $\chi_{3}^{\prime\prime}$,
$\chi_{5}^{\prime\prime}$,$\cdots$.  Figure \ref{fig:five} shows the $T$
dependence of $\chi_{3}^{\prime}$, $\chi_{5}^{\prime}$,
$\chi_{3}^{\prime\prime}$, and $\chi_{5}^{\prime\prime}$ thus obtained. 
The nonlinear susceptibility $\chi_{3}^{\prime}$ at $f$ = 1 Hz shows a
positive peak at 16.0 K, a negative local minimum at 10.2 K, and two
positive peaks at 8.8 and 8.3 K. The sign of $\chi_{3}^{\prime}$ changes
from negative to positive at 9.65 K and from positive to negative at 4.0 K
with decreasing $T$.  No anomaly is observed below 4 K. In contrast,
$\chi_{5}^{\prime}$ at $f$ = 1 Hz exhibits a negative local minimum at 16.0
K, a positive peak at 10.2 K, and two negative peaks at 9.0 K and 8.3 K.
The sign of $\chi_{5}^{\prime}$ changes from positive to negative at 4.0 K
with decreasing $T$.  We note that the $T$ dependence of
$\chi_{3}^{\prime}$ around 10 K in our system is similar to that in stage-2
CoCl$_{2}$ GIC which magnetically behave like a quasi 2D ferromagnet with
an extremely weak antiferromagnetic interplanar exchange
interaction.\cite{Suzuki2002} In stage-2 CoCl$_{2}$ GIC,
$\chi_{3}^{\prime}$ exhibits a negative local minimum at 10.5 K, becomes
positive below 10.2 K, and shows a positive peak below the upper critical
temperature $T_{cu}$.

The nonlinear susceptibility $\chi_{3}^{\prime\prime}$ shows a positive
peak at 16.2 K, a negative local minimum at 11.0 K, and a positive peak at
8.7 K. The sign of $\chi_{3}^{\prime\prime}$ changes from negative to
positive at 9.9 K and from positive to negative at 5.7 K with decreasing
$T$.  In contrast, $\chi_{5}^{\prime\prime}$ shows a negative local minimum
at 16.4 K, a positive peak at 11.0 K, and a negative peak at 9.0 K. The
sign of $\chi_{5}^{\prime\prime}$ changes from negative to positive at 6.5
K with decreasing $T$.

Here we discuss the $T$ dependence of $\chi_{3}^{\prime}$.  Basically the
singularity of $\chi_{3}^{\prime}$ could reflect the breaking of spatial
magnetic symmetry.  The sign of $\chi_{3}^{\prime}$ is negative for the PM
phase and the SG phase and positive for the FM phase.  Thus the critical
temperatures $T_{c}$ for the PM-FM transition and $T_{RSG}$ for the 
FM-RSG transition could be defined
as temperatures at which the sign of $\chi_{3}^{\prime}$ changes.  Using
this definition, in fact we find 9.65 K as $T_{c}$ and 4.0 K as $T_{RSG}$ 
at $f$ = 1 Hz for our system.  Similar behaviors have
been reported in other reentrant ferromagnets.  Sato and
Miyako\cite{Sato1981} have reported that the sign of $\chi_{3}^{\prime}$ in
(Pd$_{0.9966}$Fe$_{0.0034}$)$_{0.95}$Mn$_{0.05}$ changes from negative to
positive at $T_{c}$.  Sato et al.\cite{Sato2001} have shown that the sign
of $\chi_{3}^{\prime}$ in Ni$_{77}$$^{57}$Fe$_{1}$Mn$_{22}$ changes from
positive to negative at $T_{RSG}$: $\chi_{3}^{\prime}$ has a small negative
local minimum near $T_{RSG}$.

\subsection{\label{resultB}Linear AC susceptibility : 
$\chi^{\prime}$ and $\chi^{\prime\prime}$}

\begin{figure*}
\includegraphics[width=12.0cm]{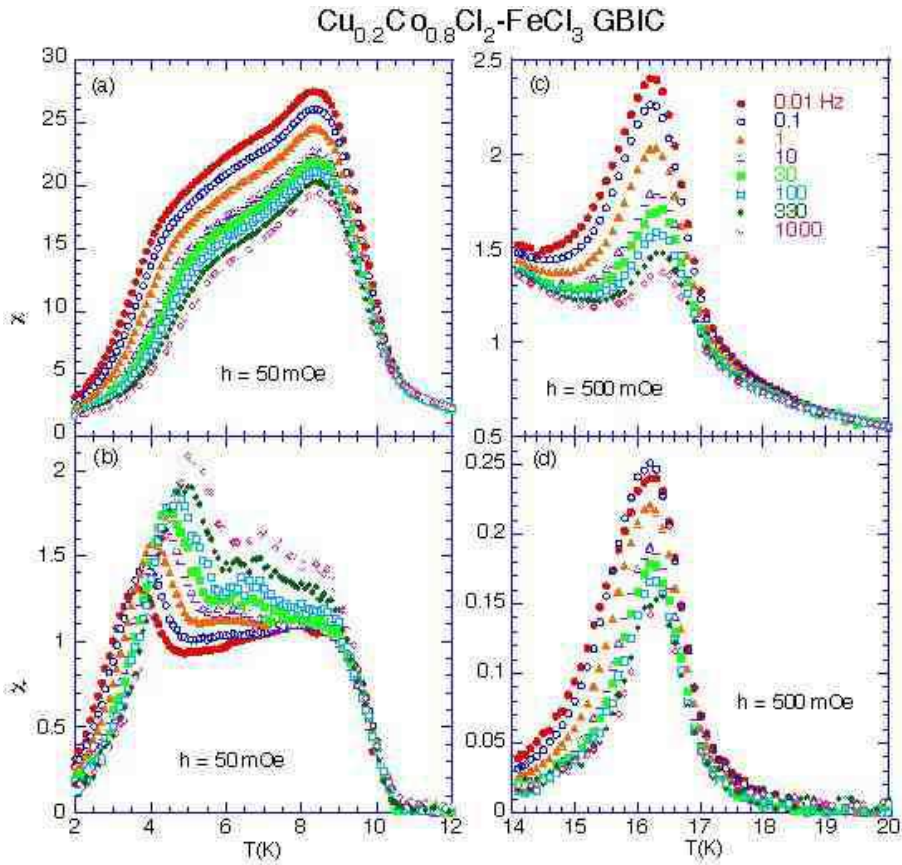}
\caption{\label{fig:six}(a), (b) $T$ dependence of $\chi^{\prime}$ (=
$\chi_{1}^{\prime}$) and $\chi^{\prime\prime}$ (=
$\chi_{1}^{\prime\prime}$) at various $f$.  $h$ = 50 mOe.  $H$ = 0.  2
$\leq T \leq$ 12 K. (c), (d) $T$ dependence of $\chi^{\prime}$ ($\approx
\chi_{1}^{\prime}$) and $\chi^{\prime\prime}$ ($\approx
\chi_{1}^{\prime\prime}$) at various $f$.  $h$ = 500 mOe.  $H$ = 0.  14
$\leq T \leq$ 20 K.}
\end{figure*}

Figures \ref{fig:six}(a) and (b) show the $T$ dependence of the linear AC
susceptibility $\chi^{\prime}$ and $\chi^{\prime\prime}$ below 12 K at $h$
= 50 mOe, where $\Theta_{1}^{\prime}/h = \chi_{1}^{\prime} = \chi^{\prime}$
and $\Theta_{1}^{\prime\prime}/h = \chi_{1}^{\prime\prime} =
\chi^{\prime\prime}$.  The absorption $\chi^{\prime\prime}$ at $f$ = 0.01 Hz
shows a relatively sharp peak at 3.69 K. This peak shifts to the high-$T$
side with increasing $f$: 5.10 K at $f$ = 1 kHz.  In contrast, the
derivative d$\chi^{\prime\prime}$/d$T$ at $f$ = 0.01 Hz shows two negative
local minima at 4.0 K corresponding to $T_{RSG}$ and 10.0 K corresponding
to $T_{c}$.  The local minimum d$\chi^{\prime\prime}$/d$T$ vs $T$ at 4.0 K
shifts to the high-$T$ side with increasing $f$, while the local minimum at
10.0 K does not shift with increasing $f$.

Figures \ref{fig:six}(c) and (d) show the $T$ dependence of
$\chi^{\prime}$ and $\chi^{\prime\prime}$ around $T_{h}$ at $h$ = 500 mOe,
where $\Theta_{1}^{\prime}/h \approx \chi^{\prime}$ and
$\Theta_{1}^{\prime\prime}/h \approx \chi^{\prime\prime}$.  The dispersion
$\chi^{\prime}$ at $f$ = 0.01 Hz shows peaks at 16.2 K and 8.35 K. The peak
at 16.2 K slightly shift to the high-$T$ side with increasing $f$: 16.4 K
at $f$ = 1 kHz.  Another peak at 8.35 K does not shift with increasing $f$. 
The dispersion $\chi^{\prime}$ at $f$ = 0.01 Hz has an inflection point at
3.5 K corresponding to the positive peak of d$\chi^{\prime}$/d$T$, and
another inflection point at 9.7 K corresponding to the negative local
minimum of d$\chi^{\prime}$/d$T$.  The inflection point at 3.5 K shift to
the high-$T$ side with increasing $f$.

\begin{figure}
\includegraphics[width=7.5cm]{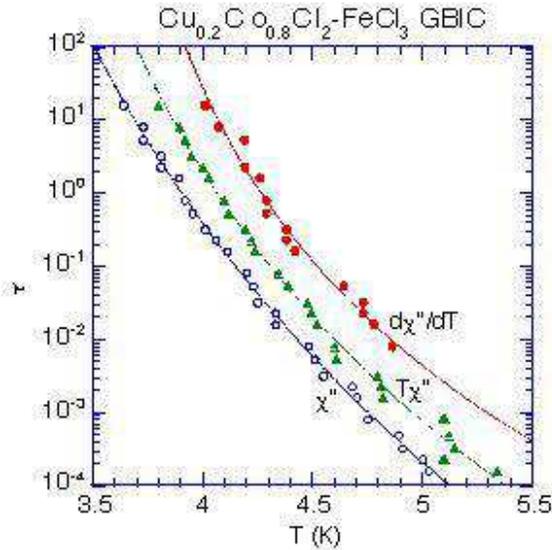}
\caption{\label{fig:seven}$T$ dependence of the relaxation time $\tau$
which is determined from the $f$ dependence of peak temperature of
$\chi^{\prime \prime}$ vs $T$, $T\chi^{\prime \prime}$ vs $T$, and
d$\chi^{\prime \prime}$/d$T$ vs $T$.  The solid lines denote the
least-squares fit of the data to Eq.(\ref{eq:three}).}
\end{figure}

Here we assume that the singular behavior of $\chi^{\prime\prime}$ around 4
K is due to the critical slowing down associated with the FM-RSG
transition.  Either the peak temperatures of $\chi^{\prime\prime}$ vs $T$
and $T\chi^{\prime\prime}$ vs $T$ or the local-minimum temperature of
d$\chi^{\prime\prime}$/d$T$ vs $T$ around 4 K coincide with a spin freezing
temperature $T_{f}$, at a relaxation time $\tau$ ($\approx 1/\omega$).  The
relaxation time $\tau$ can be described by\cite{Gunnarsson1988}
\begin{equation} 
\tau = \tau^{*}(T_{f}/T_{RSG} - 1)^{-x},
\label{eq:three} 
\end{equation} 
where 
$x = \nu z$, $z$ is the dynamic critical exponent, $\nu$ is the critical
exponent of the spin correlation length $\xi$, and $\tau^{*}$ is the
characteristic time.  In Fig.~\ref{fig:seven} we show the $T$ dependence of
$\tau$ thus obtained for d$\chi^{\prime\prime}$/d$T$,
$T\chi^{\prime\prime}$ (figure of $T\chi^{\prime\prime}$ vs $T$ is not shown), 
and $\chi^{\prime\prime}$. 
The least squares fits of the data of $\tau$ vs $T$ yield $x$ = 8.5 $\pm$
1.8, $T_{RSG}$ = 3.45 $\pm$ 0.31 K, and $\tau^{*}$ = (4.77 $\pm$ 0.10)
$\times$ 10$^{-6}$ sec for the local minimum temperature of
d$\chi^{\prime\prime}$/d$T$ vs $T$, $x$ = 12.3 $\pm$ 1.7, $T_{RSG}$ = 2.90
$\pm$ 0.26 K, and $\tau^{*}$ = (1.49 $\pm$ 0.05) $\times$ 10$^{-5}$ sec for
the peak temperature of $T\chi^{\prime\prime}$ vs $T$, and $x$ = 16.6 $\pm$
1.8, $T_{RSG}$ = 2.27 $\pm$ 0.25 K, and $\tau^{*}$ = (4.67 $\pm$ 0.10)
$\times$ 10$^{-3}$ sec for the peak temperature of $\chi^{\prime\prime}$ vs
$T$.  The value of $x$ for $\chi^{\prime\prime}$ vs $T$ and
$T\chi^{\prime\prime}$ vs $T$ is unphysically large.  In contrast, the
value of $x$ for d$\chi^{\prime\prime}$/d$T$ vs $T$ is on the same order as
that ($x$ = 7.9) rep orted by Jonason et al.\cite{Jonason1996a} for the
FM-RSG transition of
(Fe$_{0.20}$Ni$_{0.80}$)$_{75}$P$_{16}$B$_{6}$Al$_{3}$.  A relatively good
agreement between the value of $x$ for
Cu$_{0.2}$Co$_{0.8}$Cl$_{2}$-FeCl$_{3}$ GBIC and the value predicted by
Ogielski\cite{Ogielski1985} for the 3D $\pm J$ Ising SG ($x$ = 7.9 $\pm$
1.0), suggests that the FM-RSG transition in our system is dynamically
similar to an ordinary PM-SG transition.  Note that the value of $\tau^{*}$
for d$\chi^{\prime\prime}$/d$T$ vs $T$ is much larger than a typical value
of microscopic relaxation time $\tau_{0}$ (typically 10$^{-10}$ -
10$^{-12}$ sec).  Such a large value of $\tau^{*}$ has been also reported
by Kleemann et al.\cite{Kleemann2001} for
Co$_{80}$Fe$_{20}$/Al$_{2}$O$_{3}$ multilayers: $\tau^{*}$ = (6.7 $\pm$
0.4) $\times$ 10$^{-7}$ sec and $x$ = 10.0 $\pm$ 3.6.  The large $\tau^{*}$
suggests that the PM clusters play a significant role in the FM-RSG
transition.  In the random-field picture,\cite{Aeppli1983} the FM phase
consists of the FM region with a longer relaxation time and the PM clusters
with a shorter relaxation time.  On decreasing $T$ toward $T_{RSG}$, the
thermal fluctuations of the spins in the PM clusters become slower.  The
molecular field of the slow PM spins acting as random magnetic field causes
breakups of the FM network into finite-sized clusters.  It is predicted
that the following dynamic scaling equation is valid for the normal SG
phase:\cite{Geschwind1990} $T\chi^{\prime\prime} = \omega^{y}\Omega(\omega
\tau)$, where $\Omega(\omega \tau)$ is a scaling function of $\omega \tau$
and is assume to take a maximum at $\omega \tau$ = constant.  The value $y$
(= $\beta/x$) is a critical exponent, where $\beta$ is a critical exponent
of the order parameter.  The curve of $T\chi^{\prime\prime}$ vs $T$
exhibits a peak, which shifts to the high-$T$ side as $f$ increases.  The
peak height of $T\chi^{\prime\prime}$ increases with increasing $f$.  The
least squares fit of the data for the peak height of $T\chi^{\prime\prime}$
vs $f$ (for 0.01 $\leq f \leq$ 1000 Hz) to the form of ($\approx
\omega^{y}$) yields $y$ = 0.066 $\pm$ 0.001.  Then the value of $\beta$ (=
$xy$) is estimated as $\beta$ = 0.57 $\pm$ 0.10, where $x$ = 8.5 $\pm$ 1.8. 
This value of $\beta$ is similar to that ($\beta$ = 0.54) for
Fe$_{0.5}$Mn$_{0.5}$TiO$_{3}$.\cite{Gunnarsson1991} In summary, the nature
of the FM-RSG transition is similar to that of the normal PM-SG transition. 
The PM clusters for the FM-RSG transition play the same role as individual
spins for the PM-SG transition.

\begin{figure}
\includegraphics[width=7.5cm]{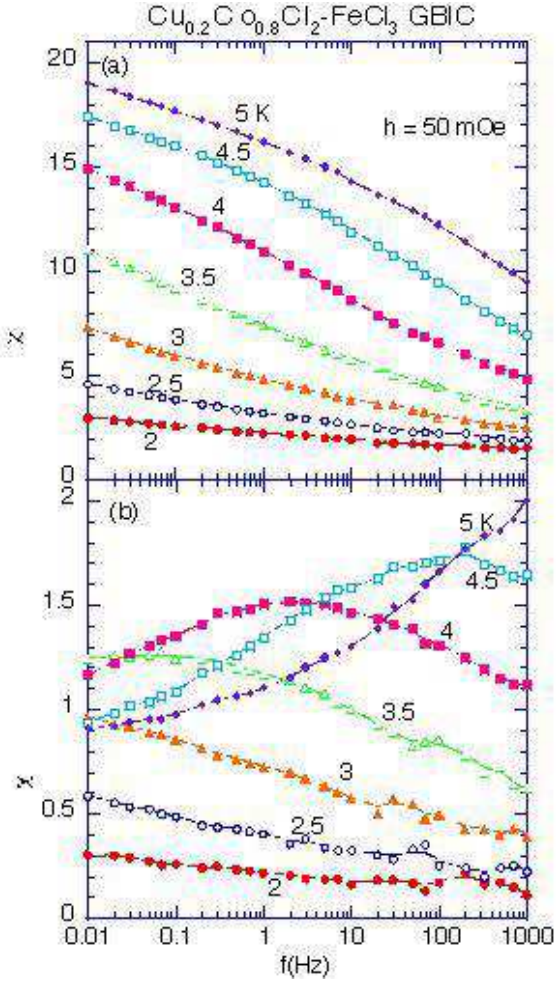}
\caption{\label{fig:eight}$f$ dependence of (a) $\chi^{\prime}$ and (b)
$\chi^{\prime\prime}$ at various $T$.  $h$ = 50 mOe.}
\end{figure}

Figures \ref{fig:eight}(a) and (b) show the $f$ dependence of
$\chi^{\prime}(T,\omega)$ and $\chi^{\prime\prime}(T,\omega)$ at various
$T$ in the vicinity of $T_{RSG}$, respectively.  The absorption
$\chi^{\prime\prime}(T,\omega)$ curves exhibit different characteristics
depending on $T$.  Above $T_{RSG}$, $\chi^{\prime\prime}(T,\omega)$ shows a
peak at a characteristic frequency, shifting to the low $f$-side as $T$
decreases.  Below $T_{RSG}$, $\chi^{\prime\prime}(T,\omega)$ shows no peak for
$f\geq$ 0.01 Hz.  It decreases with increasing $f$, following a power law
form ($\approx \omega^{-\alpha^{\prime\prime}}$).  This is in contrast to
the $f$ dependence of $\chi^{\prime\prime}$ for conventional spin glass
systems such as Fe$_{0.5}$Mn$_{0.5}$TiO$_{3}$: $\chi^{\prime\prime}$
increases with increasing $f$.\cite{Gunnarsson1988} The exponent
$\alpha^{\prime\prime}$ is weakly dependent on $T$: $\alpha^{\prime\prime}$
= 0.083 $\pm$ 0.004 at 2.5 K and $\alpha^{\prime\prime}$ = 0.079 $\pm$
0.002 at 3 K. According to the fluctuation-dissipation theorem, the
magnetic fluctuation spectrum $P(\omega)$ is related to
$\chi^{\prime\prime}(T,\omega)$ by $P(T,\omega) =
2k_{B}T\chi^{\prime\prime}(T,\omega)/\omega$.  Then $P(T,\omega)$ is
proportional to $\omega^{-1-\alpha^{\prime\prime}}$, which is similar to
1/$\omega$ character of typical spin glass.  In contrast, $\chi^{\prime}(T,
\omega)$ decreases with increasing $f$ above and below $T_{RSG}$:
$\chi^{\prime}$ exhibits a power law form ($\omega^{-\alpha^{\prime}}$). 
The exponent $\alpha^{\prime}$ is weakly dependent on $T$:
$\alpha^{\prime}$ = 0.079 $\pm$ 0.001 at $T$ = 2.5 K and $\alpha^{\prime}$
= 0.094 $\pm$ 0.001 at $T$ = 3.0 K. The value of $\alpha^{\prime}$ agrees
well with that of $\alpha^{\prime\prime}$.  Note that $\chi^{\prime\prime}$
is related to $\chi^{\prime}$ through a so called ``$\pi$/2 rule'':
$\chi^{\prime\prime} = -(\pi/2)$d$\chi^{\prime}$/d$\ln \omega$
(Kramers-Kronig relation), leading to the relation $\alpha^{\prime} =
\alpha^{\prime\prime}$.

Here we note the $f$ dependence of $\chi^{\prime\prime}$ above 5 K (which
is not shown in Fig.~\ref{fig:eight}(b)).  The absorption
$\chi^{\prime\prime}$ increases with increasing $f$ for 5 $\leq T \leq$
7.2 K. A newly small peak is added around $f$ = 2 Hz for 7.3 $\leq T \leq$
9.2 K. The absorption $\chi^{\prime\prime}$ decreases with increasing $f$
for $f \leq$ 70 Hz and increases with further increasing $f$ for 9.3 $\leq
T \leq$ 9.8 K. Above 9.9 K it decreases with increasing $f$.  We find that
$\chi^{\prime\prime}$ for 6.1 $\leq T \leq$ 7.3 K can be described by a
power law form ($\approx \omega^{\beta^{\prime\prime}}$) in the limited
frequency range (0.01 $\leq f <$ 10 Hz).  The exponent
$\beta^{\prime\prime}$ increases with increasing $T$ :
$\beta^{\prime\prime}$ = 0.04 $\pm$ 0.01 at 6.1 K, 0.081 $\pm$ 0.02 at 6.7
K, and 0.12 $\pm$ 0.01 at 7.2 K.

\subsection{\label{resultC}$\chi_{FC}$, $\chi_{ZFC}$, and $\delta \chi$ 
(= $\chi_{FC} - \chi_{ZFC}$)}

\begin{figure}
\includegraphics[width=7.0cm]{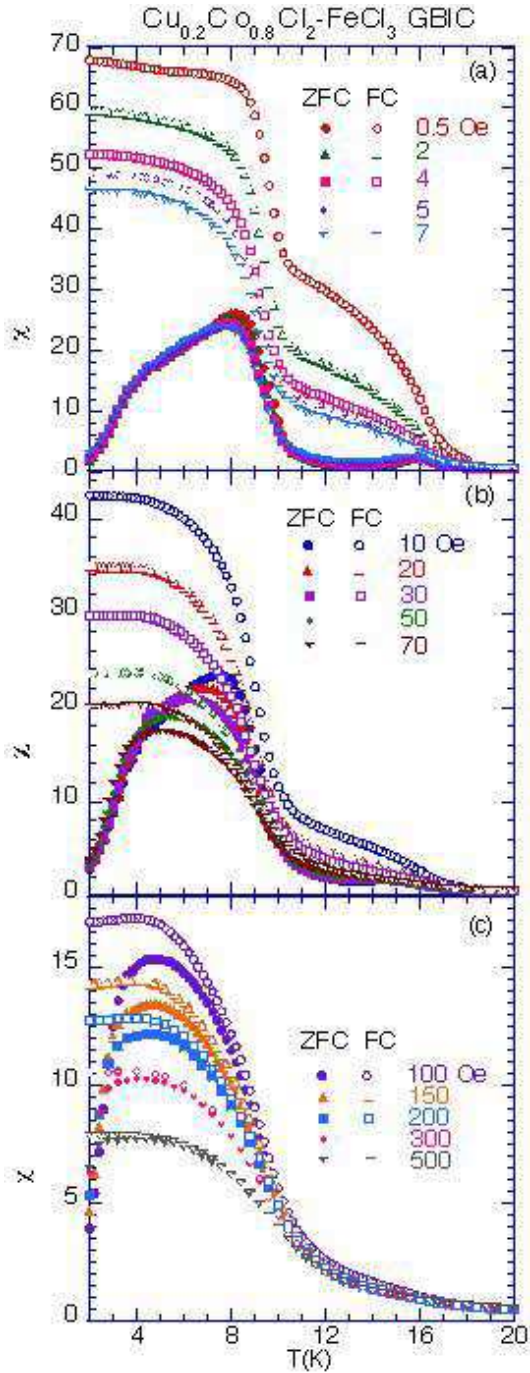}
\caption{\label{fig:nine}(a)-(c) $T$ dependence of $\chi_{ZFC}$ and
$\chi_{FC}$ at various $H$.  $H \perp c$.}
\end{figure}

Figures \ref{fig:nine} shows the $T$ dependence of
$\chi_{FC}$ and $\chi_{ZFC}$ at various $H$, where $H$ is applied along the
$c$ plane which is perpendicular to the $c$ axis.  It is strongly dependent
on $H$.  The susceptibility $\chi_{FC}$ at $H$ = 0.5 Oe decreases with
increasing $T$.  It has inflection points at $T$ = 16.4, 9.70, and 4.0
K where d$\chi_{FC}$/d$T$ exhibits negative local minima.  These
temperatures correspond to the transition temperatures $T_{h}$, $T_{c}$,
and $T_{RSG}$.  In contrast, $\chi_{ZFC}$ at $H$ = 0.5 Oe exhibits two
peaks at 16.2 and 8.2 K, and an inflection point at 3.2 K where
d$\chi_{ZFC}$/d$T$ shows a positive local maximum.  The peaks of
$\chi_{ZFC}$ at 8.2 and 16.2 K shifts to the low-$T$ side with increasing
$H$.  The deviation of $\chi_{ZFC}$ at $H$ =
0.5 Oe from $\chi_{FC}$ starts to occur at temperatures above 18 K. 
The peak of $\chi_{ZFC}$ at
$T_{h}$ is very sensitive to the application of $H$.  It shifts to the
low-$T$ side with increasing $H$ and disappears above 50 Oe.

\begin{figure}
\includegraphics[width=7.5cm]{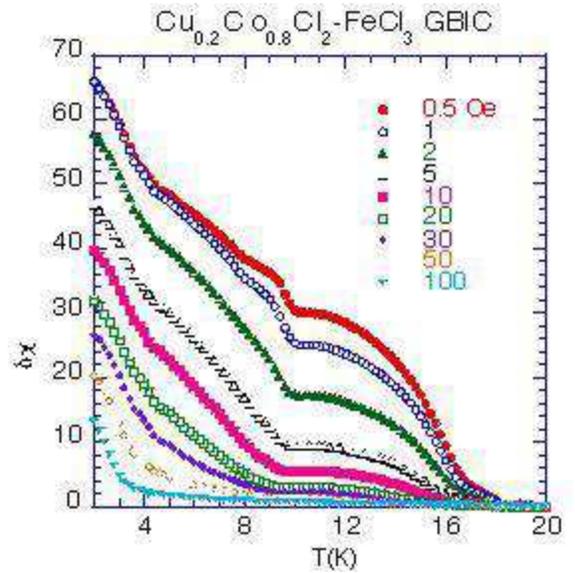}
\caption{\label{fig:eleven}$T$ dependence of $\delta \chi$ (=
$\chi_{FC} - \chi_{ZFC}$) at various $H$.  The
value of $\delta \chi$ is derived from Fig.~\ref{fig:nine}.}
\end{figure}

Figure \ref{fig:eleven} shows the $T$ dependence of the difference $\delta
\chi$ (= $\chi_{FC} - \chi_{ZFC}$) at various $H$.  The difference $\delta
\chi$ has three inflection points at $T_{h}$, $T_{c}$, and $T_{RSG}$, where
d($\delta \chi$)/d$T$ exhibits negative local minima: 16.2 K ($\approx
T_{h}$), 9.60 K ($\approx T_{c}$), and 3.40 K ($\approx T_{RSG}$).  The $T$
dependence of $\delta \chi$ shown in Fig.~\ref{fig:eleven} is very
different from standard one observed in many reentrant ferromagnets, where
$\delta \chi$ reduces to zero at $T_{RSG}$.  Typical examples of
$\chi_{ZFC}$ vs $T$ and $\chi_{FC}$ vs $T$ have been reported for
CdCr$_{2x}$In$_{2(1-x)}$S$_{4}$,\cite{Dupuis2002}
Au$_{85}$Fe$_{15}$,\cite{Razzaq1984} Ni$_{77}$Mn$_{23}$,\cite{Razzaq1987}
and (Fe$_{0.90}$Cr$_{0.05}$Ni$_{0.05}$)$_{2}$P.\cite{Srivastava1994} Both
inflection points of $\delta \chi$ at $T_{c}(H)$ and $T_{h}(H)$ become
less pronounced with increasing $H$.  Only an inflection point at
$T_{RSG}(H)$ survives for $H \geq$ 100 Oe.  Note that the inflection point
at $T_{RSG}(H)$ shifts to the low-$T$ side with increasing $H$ : 2.8 K at
$H$ = 100 Oe.  The $H$ dependence of $T_{RSG}(H)$ will be discussed in
Sec.~\ref{resultE}.

\subsection{\label{resultD}$M_{IR}^{*}$ and $M_{TR}^{*}$ in the IR and TR
states}
We have measured the magnetization $M^{*}$ in the ZFC, FC, IR (isothermal
remnant), and TR (thermoremnant) states in the case of $H$ = 5 and 15 Oe,
as a function of $T$.  This magnetization $M^{*}$ is slightly different
from usual $M$, because of different methods of measurements.  The
measurements of $M^{*}$ were carried out as follows.  First the sample was
quenched from 298 to 1.9 K at $H$ = 0.  Then $H$ (= 5 or 15 Oe) was
applied.  The measurements of $M_{ZFC}^{*}$ and $M_{IR}^{*}$ were done with
increasing $T$ from 1.9 to 20 K. At each $T$, $M_{ZFC}^{*}$ was measured at
the same $H$ and then $M_{IR}^{*}$ was measured 100 sec later after the
field was changed from $H$ to 0 Oe.  Second, the sample was annealed at 100
K for 1200 sec at $H$.  The measurements of $M_{FC}^{*}$ and $M_{TR}^{*}$
were done with decreasing $T$ from 20 to 1.9 K. At each $T$, $M_{FC}^{*}$
was measured at $H$ and then $M_{TR}^{*}$ was measured 100 sec later after
the field was changed from $H$ to 0 Oe.

\begin{figure*}
\includegraphics[width=12.0cm]{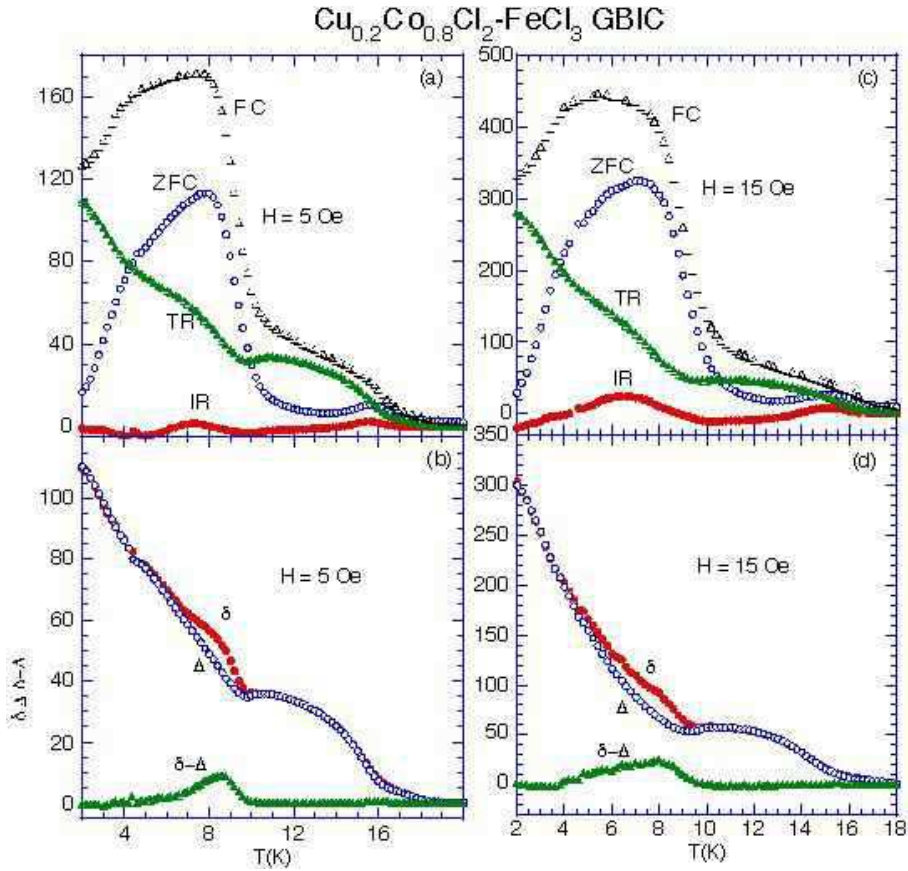}
\caption{\label{fig:twelve}$T$ dependence of $M_{ZFC}^{*}$, $M_{FC}^{*}$,
$M_{IR}^{*}$, $M_{TR}^{*}$, $\delta M^{*}$ (= $M_{FC}^{*} - M_{ZFC}^{*}$)
and $\Delta M^{*}$ (= $M_{TR}^{*} - M_{IR}^{*}$).  $H \perp c$.  (a), (b)
$H$ = 5 Oe and (c), (d) $H$ = 15 Oe.  The definition of $M^{*}$ for each
state is given in the text.}
\end{figure*}

Figure \ref{fig:twelve} shows the $T$ dependence of $M_{ZFC}^{*}$,
$M_{IR}^{*}$, $M_{FC}^{*}$, and $M_{TR}^{*}$, $\delta M^{*}$ (=
$M_{FC}^{*} - M_{ZFC}^{*}$), $\Delta M^{*}$ (= $M_{TR}^{*} - M_{IR}^{*}$)
in the case of $H$ = 5 and 15 Oe.  Note that the $T$ dependence of
$M_{FC}^{*}$ is not exactly the same as that of $\chi_{FC}$, because of the
difference in the method of field cooling.  While $\chi_{FC}$ shown in
Fig.~\ref{fig:nine}(a) increases with decreasing $T$, $M_{FC}^{*}$ shows a
peak at 7.5 K between $T_{c}$ and $T_{RSG}$ and an inflection point at
$T_{RSG}$ where d$M_{FC}^{*}$/d$T$ exhibits a positive local maximum.  This
result is indicative of a non-uniformity in the FM phase: frustrated spins
coexists with ferromagnetically aligned spins.

The $T$ dependence of $M_{TR}^{*}$ is exactly the same as that of $\delta
M^{*}$.  The magnetization $M_{TR}^{*}$ (measured at $H$ = 0) in the case
of $H$ = 5 Oe shows a broad peak around 10.6 K just above $T_{c}$, and
reduces to zero at $T_{h}$.  In contrast, $M_{IR}^{*}$ is much smaller than
$M_{TR}^{*}$.  The difference $\delta M^{*}$ is different from $\Delta
M^{*}$ between $T_{RSG}$ and $T_{c}$.  In fact, the difference ($\delta
M^{*} -\Delta M^{*}$) is larger than zero only between $T_{RSG}$ and
$T_{c}$ and near $T_{h}$.

As far as we know, there has been only one report on $M_{TR}$ (measured by
a conventional method) in a reentrant ferromagnet
(Fe$_{0.90}$Cr$_{0.05}$Ni$_{0.05}$)$_{2}$P.\cite{Srivastava1994} 
$M_{TR}$ was measured as $T$ increases after cooling down the system to the
lowest $T$ in the presence of $H$ (FC cooling), annealing at $T$ at a
waiting time $t_{w}$, and then reducing $H$ to zero.  The magnetization
$M_{TR}$ shows a local minimum at $T_{RSG}$ and a broad peak between
$T_{RSG}$ and $T_{c}$.  It reduces to zero at $T_{c}$ with increasing $T$.

\subsection{\label{resultE}$H$-$T$ phase diagram}

\begin{figure*}
\includegraphics[width=12.0cm]{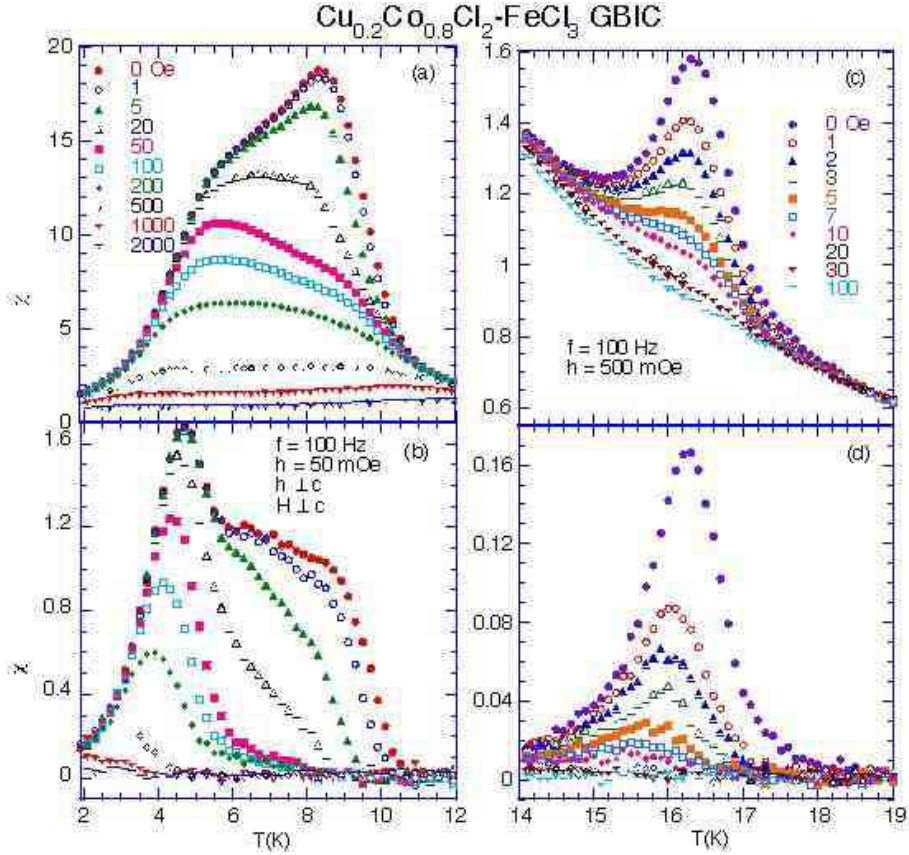}
\caption{\label{fig:thirteen}$T$ dependence of (a) $\chi^{\prime}$ and (b)
$\chi^{\prime\prime}$ at various $H$.  $H$ is applied along the $c$ plane
perpendicular to the graphene plane ($H \perp c$).  $f$ = 100 Hz.  $h$ = 50
mOe.  $h \perp c$.  1.9 $\leq T \leq$ 12 K. $T$ dependence of (c)
$\chi^{\prime}$ and (d) $\chi^{\prime\prime}$ at various $H$.  $H \perp c$. 
$f$ = 100 Hz.  $h$ = 500 mOe.  $h \perp c$.  14 $\leq T \leq$ 19 K.}
\end{figure*}

\begin{figure}
\includegraphics[width=7.5cm]{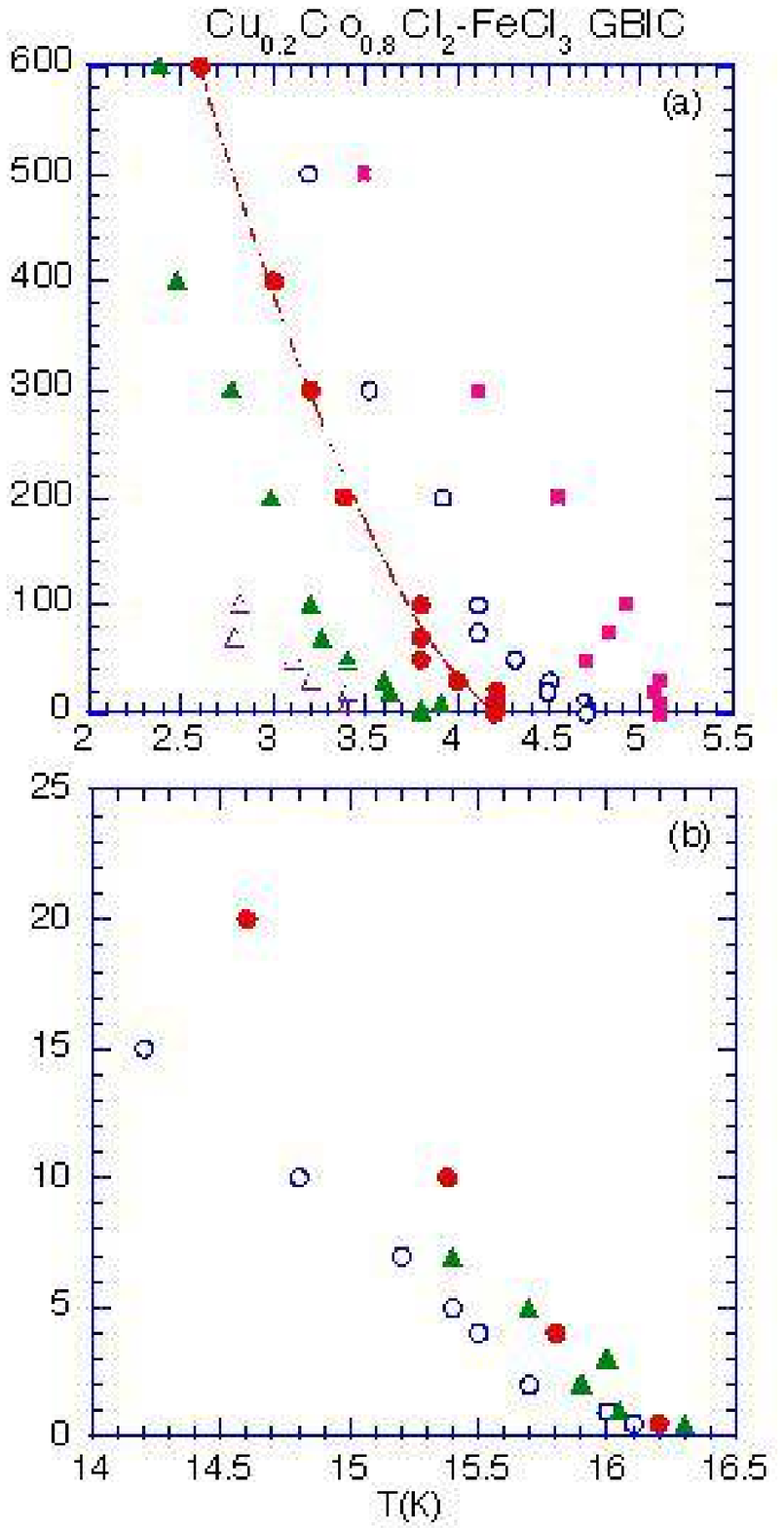}
\caption{\label{fig:fifteen}(a) $H$-$T$ phase diagram near $T_{RSG}$: for
each $H$ the local-minimum temperatures of d$\chi^{\prime\prime}$/d$T$ vs
$T$ at \textit{f} = 0.1 Hz ({\Large $\bullet$}) and at 100 Hz
($\blacksquare$), and d$\delta \chi$/d$T$ vs $T$ ($\triangle$), and the
peak temperatures of $\chi^{\prime\prime}$ vs $T$ at 0.1 Hz
($\blacktriangle$) and 100 Hz ({\Large $\circ$}).  
The solid line denotes the least squares fitting curve (see the text 
for detail).
(b) $H$-$T$ phase
diagram near $T_{h}$: for each $H$ the local-minimum temperature of
d$\delta \chi$/d$T$ vs $T$ ({\Large $\circ$}), and the peak temperatures
of $\chi_{ZFC}$ vs $T$ ({\Large $\bullet$}) and $\chi^{\prime\prime}$ vs
$T$ at 100 Hz ($\blacktriangle$).}
\end{figure}

We have measured the $T$ dependence of $\chi^{\prime}$ and 
$\chi^{\prime\prime}$ at various $H$ for $f = 100$ Hz and 0.1 Hz, 
where $H$ ($0 < H \leq 2$ kOe) is applied along the $c$ plane 
perpendicular to the $c$-axis. 
Figures \ref{fig:thirteen} shows the $T$ dependence
of $\chi^{\prime}$ and $\chi^{\prime\prime}$ at various $H$ for $f$ = 
100 Hz.
The AC field $h$ (= 50 mOe)
was used for 1.9 $\leq T \leq$ 12 K and a larger $h$ (= 500 mOe) was used
for 14 $\leq T \leq$ 19 K. The peak of $\chi^{\prime}$ and the shoulder of
$\chi^{\prime\prime}$ around $T_{c}$ disappears for $H \geq$ 50 Oe, and the
peaks of $\chi^{\prime}$ and $\chi^{\prime\prime}$ around $T_{h}$
disappears for $H \geq$ 7 Oe.  The peak of $\chi^{\prime\prime}$ around
$T_{RSG}$ shifts to the low $T$-side with increasing $H$ for 0 $\leq H
\leq$ 1 kOe.  The peak of $\chi^{\prime\prime}$ around $T_{h}$ also shifts
to the low $T$-side with increasing $H$ for 0 $\leq H \leq$ 7 Oe.  In
Figs.~\ref{fig:fifteen}(a) and (b) we show the $H$-$T$ diagrams around
$T_{RSG}$ and $T_{h}$, respectively.  Here the temperature (denoted 
as $T_{f}(H)$) of negative local minimum of d$\chi^{\prime\prime}$/d$T$ vs
$T$ at 0.1 Hz (data are not shown) is plotted as a function of $H$.  
For comparison, we also
show the $H$ dependence of the negative local minimum temperatures of
d$\delta \chi$/d$T$ vs $T$ and the peak temperatures of
$\chi^{\prime\prime}$ vs $T$ at $f$ = 0.1 and 100 Hz.  These lines are away
from the line $T_{f}(H)$.  The least squares fit of the data of the line
$T_{f}(H)$ for 0 $\leq H \leq$ 600 Oe to
\begin{equation} 
H = H_{0}^{*}[1-T_{f}(H)/T_{g}^{*}]^{p}, 
\label{eq:four} 
\end{equation} 
yields parameters $p$ = 1.55 $\pm$ 0.13, $T_{g}^{*}$ = 4.26 $\pm$ 0.11 K, and
$H_{0}^{*}$ = (2.6 $\pm$ 0.3) kOe.  Note that the value of $T_{g}^{*}$ (=
4.26 K) is larger than that of $T_{RSG}$ (= 3.45 $\pm$ 0.31 K).  The value
of exponent $p$ is close to that ($p$ = 1.50) for the de Almeida-Thouless
(AT) line.\cite{Almeida1978} In the mean field picture, the AT line
separates the replica-symmetry (FM) phase from the
replica-symmetry-breaking (RSB) phase in the ($H$,$T$) line.

\subsection{\label{resultF}Aging: $M_{ZFC}(t,T)$,
$\chi^{\prime\prime}(t,\omega)$, and $\chi^{\prime}(t,\omega)$}

\begin{figure}
\includegraphics[width=7.5cm]{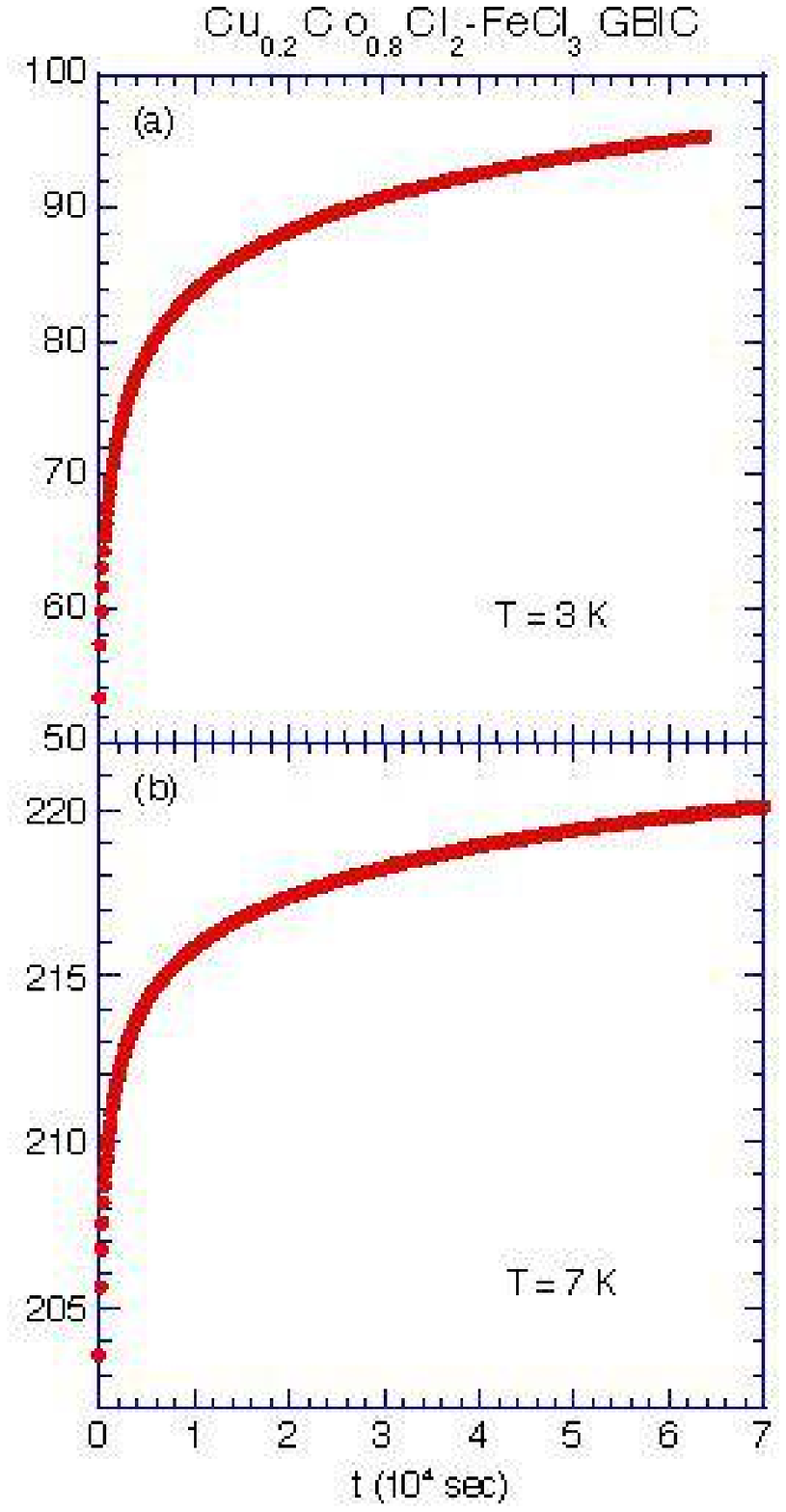}
\caption{\label{fig:sixteen}$t$ dependence of $M_{ZFC}$ at $T$ = 3 
and 7 K. 
$t_{w}$ = 2.0 $\times$ 10$^{3}$ sec.  $H$ = 10 Oe.  $H \perp c$.  See the
detail of the measurement and the definition of $t$ = 0 in the text.}
\end{figure}

\begin{figure}
\includegraphics[width=7.5cm]{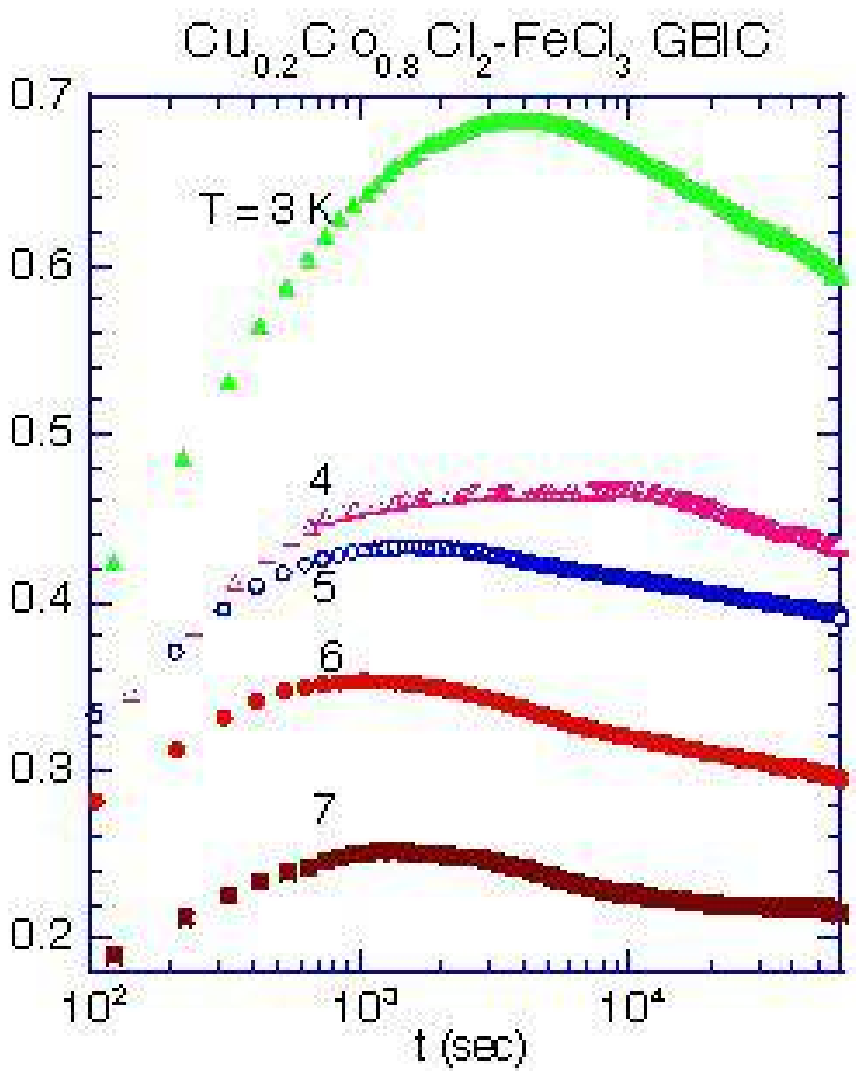}
\caption{\label{fig:seventeen}$t$ dependence of $S(t)$ [$=
(1/H)\text{d}M_{ZFC}/\text{d}\ln t$] at various $T$.  $t_{w} = 2.0 
\times 10^{3}$ sec.}
\end{figure}

In order to reveal a possible aging phenomenon in the RSG phase and FM
phase, we have studied time ($t$) dependence of the zero-field cooled
magnetization $M_{ZFC}$.  First the system was cooled from 100 K to $T$ in
the absence of $H$, where the remnant magnetic field is less than 3 mOe. 
The system was kept at $T$ for a wait time $t_{w}$ (= 2.0 $\times$ 10$^{3}$
sec).  A DC magnetic field ($H$ = 10 Oe) was applied at $t$ = 0.  The
magnetization ($M_{ZFC}$) was measured as a function of $t$ elapsed after
the field application.  Figure \ref{fig:sixteen} shows the $t$ dependence
of $M_{ZFC}(t)$ at $T$ = 3 and 7 K.  The corresponding relaxation rate $S(t)$ [
$= (1/H)\text{d}M_{ZFC}(t)/\text{d}\ln t$] is shown in Fig.~\ref{fig:seventeen} as
a function of $t$ for each $T$.  The relaxation rate $S(t)$ at $T$ = 3 K
has a peak around $t_{p}$ = 4 $\times$ 10$^{3}$ sec which is larger than
$t_{w}$.  Note that the inflection point of $M_{ZFC}$ corresponds to the
peak of $S(t)$.  At $T$ = 4 K, $S(t)$ shows a very flat plateau between 3.2
$\times$ 10$^{3}$ and 10$^{4}$ sec, in addition to a small peak at $t_{p}$ =
1.1 $\times$ 10$^{4}$ sec.  Such an aging behavior does not end at
$T_{RSG}$, but sustains into the FM phase.  The relaxation rate $S(t)$ has
a peak at a time shorter than $t_{w}$ for 5 $\leq T \leq$ 7 K: $t_{p}$ =
1.6 $\times$ 10$^{3}$ sec at $T$ = 5 K, 1 $\times$ 10$^{3}$ sec at 6 K, and
1.3 $\times$ 10$^{3}$ sec at 7 K. The peak time $t_{p}$ is equal to is 2.1
$\times$ 10$^{3}$ sec at 8 K and 1.8 $\times$ 10$^{3}$ sec at 9 K, which
are close to $t_{w}$.  Similar behavior is observed in the FM phase of
(Fe$_{0.20}$Ni$_{0.80}$)$_{75}$P$_{16}$B$_{6}$Al$_{3}$ by Jonason et
al.\cite{Jonason1996a,Jonason1996b}: $S(t)$ has a peak around $t_{w}$ (=
100 - 10$^{4}$ sec).

\begin{table}
\caption{\label{table:one}Least squares fitting parameters of $M_{ZFC}(t)$
to the power law form given by Eq.(\ref{eq:six}).  $H$ = 10 Oe.  $t_{w}$ =
2.0 $\times$ 10$^{3}$ sec.  The values of $M_{FC}$ and $M_{ZFC}$ are obtained
from Fig.~\ref{fig:nine}(b) at $H$ = 10 Oe.  $M_{1}^{\prime}$,
$M_{2}^{\prime}$, $M_{FC}$ and $M_{ZFC}$ are in the units of emu/av mol.}
\begin{ruledtabular}
\begin{tabular}{llllll}
    $T(K)$ & $a$ & $M^{\prime}_{1}$ & $M^{\prime}_{2}$ & $M_{FC}$ & 
    $M_{ZFC}$\\
5 & 0.015 $\pm$ 0.001 & 461.6 & 290.3 & 414.4 & 187.1\\
6 & 0.035 $\pm$ 0.001 & 296.0 & 99.0 & 399.8 & 214.6\\
7 & 0.030 $\pm$ 0.001 & 292.8 & 82.4 & 374.1 & 231.1\\
8 & 0.059 $\pm$ 0.001 & 232.4 & 26.6 & 327.3 & 229.4\\
9 & 0.145 $\pm$ 0.005 & 140.6 & 6.0 & 221.5 & 155.6\\
\end{tabular}
\end{ruledtabular}
\end{table}

The relaxation rate $S(t)$ clearly shows a crossover between two asymptotic
relaxation regimes: the peak time $t_{p}$ in the RSG phase is much longer
than that in the FM phase under the same value of $t_{w}$.  The FM phase of
our system is chaotic in a similar way as the RSG phase.  This is in
contrast to the nonfrustrated nature of regular FM phase.  The relaxation
mechanism in the FM phase is different from that in the RSG phase in our
system.  The relaxation of $M_{ZFC}(t)$ at $T$ = 3 and 4 K is described by
the superposition of a stretched exponential and a
constant:\cite{Roshko1994}
\begin{equation} 
M_{ZFC}(t) = M_{1}-M_{2}\exp[-(t/\tau_{M})^{1-n}],
\label{eq:five} 
\end{equation} 
where $\tau_{M}$ is a relaxation time, $M_{1}$ and $M_{2}$ are constants, the
exponent $n$ = 0 corresponds to the Debye, single time-constant exponential
relaxation and $n$ = 1 corresponds to $t$-independent $M_{ZFC}$.  The least
squares fit of the data to Eq.(\ref{eq:five}) yields $n$ = 0.761 $\pm$
0.002, $\tau_{M}$ = (6.20 $\pm$ 0.05) $\times$ 10$^{3}$ sec, $M_{1}$ = 108.72
$\pm$ 0.17 (emu/av mol), $M_{2}$ = 76.93 $\pm$ 0.53 (emu/av mol) for $T$ =
3 K, and $n$ = 0.819 $\pm$ 0.001, $\tau_{M}$ = (6.49 $\pm$ 0.04) $\times$
10$^{3}$ sec, $M_{1}$ = 171.53 $\pm$ 0.11 (emu/av mol), $M_{2}$ = 70.28
$\pm$ 0.33 (emu/av mol) for $T$ = 4 K. In contrast, the relaxation of
$M_{ZFC}(t)$ for 5 $\leq T \leq$ 9 K can be well described by a simple
power law form and a constant:\cite{Roshko1994,Ito1986}
\begin{equation} 
M_{ZFC}(t) = M_{1}^{\prime}- M_{2}^{\prime}t^{-a},
\label{eq:six} 
\end{equation} 
where $a$ is the exponent, and $M_{1}^{\prime}$ and $M_{2}^{\prime}$ are
constant magnetizations.  The magnetization $M_{1}^{\prime}$ is the
saturation value which $M_{ZFC}$ reaches in the limit of $t \rightarrow
\infty$.  The least squares fit of the data to Eq.(\ref{eq:six}) yields the
parameters listed in Table \ref{table:one}.  The exponent $a$ is roughly on
the same order as that reported by Li et al.\cite{Roshko1994} for a
reentrant ferromagnet (Fe$_{0.65}$Ni$_{0.35}$)$_{0.882}$Mn$_{0.118}$ ($a$ =
0.088 - 0.060 for $T_{c} < T < T_{RSG}$).  However, the value of $a$ in our
system tends to increase with increasing $T$ between $T_{RSG}$ and $T_{c}$,
while the value of $a$ in (Fe$_{0.65}$Ni$_{0.35}$)$_{0.882}$Mn$_{0.118}$
decreases with increasing $T$.  We note that the $T$ dependence of $a$ in
our system near $T_{c}$ is similar to that in Fe$_{0.5}$Mn$_{0.5}$TiO$_{3}$
near $T_{SG}$,\cite{Ito1986} in spite of the fact that
Fe$_{0.5}$Mn$_{0.5}$TiO$_{3}$ is a pure spin glass and undergoes a
transition between the SG phase and the PM phase at $T_{SG}$.  The values
of $M_{FC}$ and $M_{ZFC}$ measured at $H$ = 10 Oe (see
Fig.~\ref{fig:nine}(b)) are also listed in Table \ref{table:one}.  The
value of $M_{1}^{\prime}$ is on the same order as that of $M_{FC}$ and much
larger than that of $M_{ZFC}$, where $M_{FC}$ is close to thermodynamic
equilibrium value.  The prefactor $M_{2}^{\prime}$ decreases with
increasing $T$ and tends to reduce to zero around $T_{c}$.

\begin{figure}
\includegraphics[width=8.5cm]{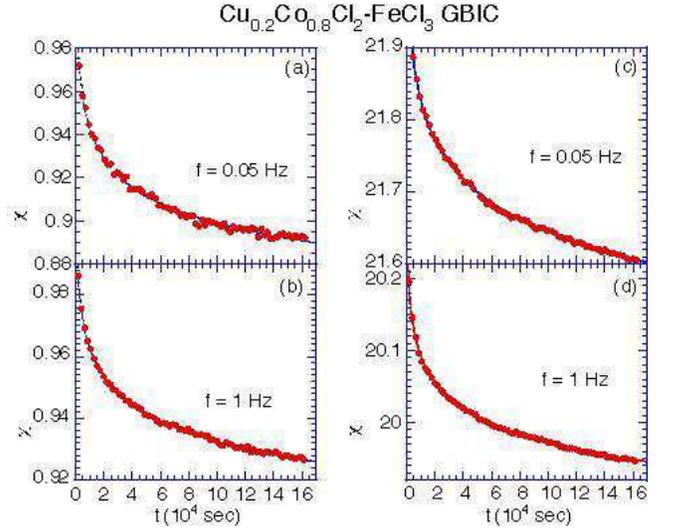}
\caption{\label{fig:eighteen}$t$ dependence of (a) and (b)
$\chi^{\prime\prime}$, and (c) and (d) $\chi^{\prime}$ at $T$ = 7 K for $f$ =
0.05 and 1 Hz, where $t$ is the time taken after the sample was
quenched from 100 K to $T$ = 7 K. $H$ = 0.  The solid lines denote the
least squares fitting curves (see the text for detail).}
\end{figure}

\begin{table}
\caption{\label{table:two}Exponents $b^{\prime\prime}$ and $b^{\prime}$
determined from the least squares fits of $\chi^{\prime\prime}(t,\omega)$
and $\chi^{\prime}(t,\omega)$ at $f$ to the power law form by given by
Eq.(\ref{eq:seven}) for $\chi^{\prime\prime}( t,\omega)$ and the
corresponding equation for $\chi^{\prime}(t,\omega)$.  $T$ = 7 and 8.5 K.}
\begin{ruledtabular}
\begin{tabular}{llll}
    $T(K)$ & $f(Hz)$ & $b^{\prime\prime}$ & $b^{\prime}$\\
7 & 0.05 & 0.074 $\pm$ 0.016 & \_\\
7 & 0.1 & 0.045 $\pm$ 0.013 & 0.012 $\pm$ 0.005\\
7 & 0.5 & 0.042 $\pm$ 0.008 & 0.045 $\pm$ 0.005\\
7 & 1 & 0.029 $\pm$ 0.006 & 0.046 $\pm$ 0.006\\
7 & 5 & 0.015 $\pm$ 0.025 & 0.070 $\pm$ 0.016\\
8.5 & 0.05 & 0.147 $\pm$ 0.067 & \_\\
8.5 & 0.5 & 0.065 $\pm$ 0.028 & \_\\
8.5 & 1 & 0.046 $\pm$ 0.022 & 0.038 $\pm$ 0.011\\
\end{tabular}
\end{ruledtabular}
\end{table}

We have measured the $t$ dependence of $\chi^{\prime\prime}$ at $T$ = 7 and
8.5 K, where $H$ = 0.  The system was quenched from 100 K to $T$ at time
(age) zero.  Both $\chi^{\prime}$ and $\chi^{\prime\prime}$ were measured
simultaneously as a function of time $t$ at constant $T$.  Each point
consists in the successive measurements at various frequencies ($0.05 
\leq f \leq 1$ Hz).  Figure
\ref{fig:eighteen} shows the $t$ dependence of $\chi^{\prime\prime}$ and
$\chi^{\prime}$ at $T$ = 7 K for $f$ = 0.05 and 1 Hz,
respectively.  The absorption $\chi^{\prime\prime}$ decreases with
increasing $t$ and is well described by a power-law decay
\begin{equation} 
\chi^{\prime\prime}(t,\omega) = \chi^{\prime\prime}_{0}(\omega) 
+A_{0}^{\prime\prime}(\omega) t^{-b^{\prime\prime}}, 
\label{eq:seven} 
\end{equation} 
where $b^{\prime\prime}$ is an exponent, and
$\chi^{\prime\prime}_{0}(\omega)$ and $A_{0}^{\prime\prime}(\omega)$ are
$t$-independent constants.  In the limit of $t \rightarrow \infty$,
$\chi^{\prime\prime}(t,\omega)$ tends to $\chi^{\prime\prime}_{0}(\omega)$. 
The least squares fit of the data of $\chi^{\prime\prime}(t,\omega)$ at $T$
= 7 K to Eq.(\ref{eq:seven}) yields parameters listed in Table
\ref{table:two}.  The exponent $b^{\prime\prime}$, which is dependent on
$f$, is smaller than that of the FM phase of reentrant ferromagnet
CdCr$_{1.8}$In$_{0.2}$S$_{4}$ ($b^{\prime\prime} \approx$
0.2)\cite{Dupuis2002} and the SG phase of pure spin glass
Fe$_{0.5}$Mn$_{0.5}$TiO$_{3}$ ($b^{\prime\prime} \approx$ 0.14 $\pm$
0.03).\cite{Dupuis2001} The value of $\chi^{\prime\prime}_{0}$ tends to
decrease with increasing $f$.  In contrast, the value of $A^{\prime\prime}$
tends to increase with increasing $f$.  It follows that the second term of
Eq.(\ref{eq:seven}) cannot be described by a power form $(\omega
t)^{-b^{\prime\prime}}$, suggesting no $\omega t$-scaling law in the form
of $\chi^{\prime\prime} \approx (\omega t)^{-b^{\prime\prime}}$.  This is
in contrast to the $\omega t$-scaling of $\chi^{\prime\prime}$ in the FM
phase of reentrant ferromagnet
CdCr$_{1.8}$In$_{0.2}$S$_{4}$.\cite{Dupuis2002} The $t$ dependence of
$\chi^{\prime}$ can be also described by the power law form ($\approx
t^{-b^{\prime}}$) which is similar to Eq.(\ref{eq:seven}).  The value of
$b^{\prime}$ at $T$ = 7 and 8.5 K which is listed in Table \ref{table:two},
is on the same order as that of $b^{\prime\prime}$.  Note that similar
aging behavior is also observed both in $\chi^{\prime\prime}$ and
$\chi^{\prime}$ below $T_{RSG}$.  The change of $\chi^{\prime\prime}$ and
$\chi^{\prime}$ with $t$ below $T_{RSG}$ is not so prominent compared to
that above $T_{RSG}$, partly because of relatively small magnitude of
$\chi^{\prime\prime}$ and $\chi^{\prime}$ below $T_{RSG}$.

\section{\label{dis}DISCUSSION}
The RSG phase below $T_{RSG}$ is not a mixed phase but a normal SG phase. 
The static and dynamic behaviors of the RSG phase are characterized by that
of the normal SG phase: a critical exponent $\beta$ = 0.57 $\pm$ 0.10, a
dynamic critical exponent $x$ = 8.5 $\pm$ 1.8, and an exponent $p$ (= 1.55
$\pm$ 0.13) for the AT line.\cite{Almeida1978} The aging phenomena are
observed.  The relaxation of $M_{ZFC}(t)$ obeys a stretched exponential
law, which is usually used in analysis of the dynamics of the normal 
SG phase.  No appreciable nonlinear magnetic susceptibility is observed
below $T_{RSG}$.  These results suggest that the long-range ferromagnetic
correlation completely disappears in the RSG phase.

In contrast, the FM phase of our system is very different from that of
regular FM phase.  The chaotic behavior observed in the FM phase is 
rather similar to that
in the RSG phase.  The prominent nonlinear susceptibility of the
FM phase arises mainly from the unfrustrated ferromagnetically ordered
spins (the FM network).  The aging phenomena are also observed in the FM
phase.  A dynamic crossover of the relaxation of$M_{ZFC}(t)$ is observed
around $T_{RSG}$.  Above $T_{RSG}$ the relaxation of $M_{ZFC}(t)$ obeys a
weak power law.  These results can be well explained in terms of the
phenomenological random-field picture proposed by Aeppli et
al.\cite{Aeppli1983} (see Sec.~\ref{intro}).  The FM phase consists of the
FM network (with slow dynamics) surrounded by frustrated spins (the PM
clusters with fast dynamics).  Such a nonuniformity gives rise to the
chaotic nature of the FM phase.  On approaching $T_{RSG}$ from the high-$T$
side, the thermal fluctuations of the spins in the PM clusters become so
slow that the slow dynamics of the FM network is significantly influenced
by the dynamics of the PM clusters.  The coupling between the FM network
and the PM clusters becomes important.  The molecular fields from the slow
PM spins act a random magnetic field.  This causes breakups of the FM
network in to finite-sized clusters.  Below $T_{RSG}$, the system becomes
into the RSG phase.

Next we discuss the nature of the aging phenomena in the FM phase.  The
features of the aging phenomena are summarized as follows.  (i)
$M_{ZFC}(t)$ has a power law form ($M_{ZFC} \approx -t^{-a}$) for 5 $\leq T
\leq$ 9 K. The value of $a$ is listed in Table \ref{table:one}.  (ii) Both
$\chi^{\prime\prime}(t,\omega)$ and $\chi^{\prime}(t,\omega)$ at $T$ = 7
and 8.5 K have power law forms ($\chi^{\prime\prime} \approx
t^{-b^{\prime\prime}}$ and $\chi^{\prime} \approx t^{-b^{\prime}}$) at fixed
$f$ (=$\omega/2\pi$).  The values of $b^{\prime}$ and $b^{\prime\prime}$
are listed in Table \ref{table:two}.  (iii) $\chi^{\prime\prime}(\omega)$
for 6.1 $\leq T \leq$ 7.3 K has a power law form ($\chi^{\prime\prime} \approx
\omega^{\beta^{\prime\prime}}$), where $\beta^{\prime\prime}$ = 0.04 $\pm$
0.01 at 6.1 K and 0.12 $\pm$ 0.01 at 7.2 K. The
increase of $\chi^{\prime\prime}(\omega)$ with increasing $\omega$ is
similar to that reported in conventional SG's such as
Fe$_{0.5}$Mn$_{0.5}$TiO$_{3}$.\cite{Gunnarsson1988}

As is listed in Table \ref{table:one}, the exponent $a(T)$ increases
exponentially with increasing $T$ and is described by $a(T) = 0.23
\exp[-(1-T/T_{c})/0.148]$ with $T_{c}$ = 9.7 K. Similar result on $a(T)$
has been reported by Ito et al.\cite{Ito1986} for the SG phase of
Fe$_{0.5}$Mn$_{0.5}$TiO$_{3}$: $a(T) = 1.6 \exp\{-[1-T/T_{SG}(H)]/0.17\}$ with
$T_{SG}(H)$ = 17.6 K at $H$ = 3.2 kOe.  From Monte-Carlo simulations on the
3D $\pm J$ Ising spin glass model, Ogielski\cite{Ogielski1985} has
discussed the $t$ dependence of the order parameter $q(t)$ below the spin
freezing temperature $T_{SG}$.  The order parameter $q(t)$ is described by
a power law form [$q(t) \approx t^{-a(T)}$], where $a(T)$ is well fitted by
$a(T) \approx 0.07 \exp[-(1-T/T_{SG})/0.28]$:\cite{Ito1986} $a$ = 0.07 at
$T$ = $T_{SG}$.  The similarity between our result and the computer
simulation is remarkable.  This implies that the growth of $M_{ZFC}$ in the
FM phase (but not in the RSG phase) is essentially the same as that 
in the SG phase of spin glass systems.

What is the relation between the exponent $a$ for $M_{ZFC}(t)$ and the
exponent ($b^{\prime}$, $b^{\prime\prime}$, $\beta^{\prime\prime}$) for
$\chi^{\prime}$ and $\chi^{\prime\prime}$?  According to Lundgren et
al.,\cite{Lundgren1990,Lundgren1982} $M_{ZFC}(t)$ is related to the linear
AC susceptibility $\chi^{\prime\prime}(\omega)$ and $\chi^{\prime}(\omega)$
through the following relations
\begin{equation} 
(1/H)\text{d}M_{ZFC}/\text{d}\ln t = 2\chi^{\prime\prime}(\omega)/\pi, 
\hspace{5mm} (t=1/\omega),
\label{eq:eight} 
\end{equation} 
and
\begin{equation} 
1- q(t) = (1/H)M_{ZFC}(t) = \chi^{\prime}(\omega), 
\hspace{5mm} (t= 1/\omega).
\label{eq:nine} 
\end{equation} 
When $M_{ZFC}(t)$ is described by a power law form given by
Eq.(\ref{eq:six}), $\chi^{\prime\prime}(\omega)$ and
$\chi^{\prime}(\omega)$ can be estimated as $\chi^{\prime\prime}(\omega)
\approx \omega^{a}$ (or $t^{-a}$) and $\chi^{\prime}(\omega) \approx
\omega^{a}$ (or $t^{-a}$), respectively, leading to a prediction that the
exponents $\beta^{\prime\prime}$, $b^{\prime}$, and $b^{\prime\prime}$ are
essentially the same as the exponent $a$.  Experimentally we find $a$ =
0.059 $\pm$ 0.001 ($T$ = 8 K), $b^{\prime\prime}$ = 0.074 $\pm$ 0.016 ($T$
= 7 K and $f$ = 0.05 Hz), and $b^{\prime}$ = 0.046 $\pm$ 0.006 ($T$ = 7 K,
$f$ = 1 Hz), and $\beta^{\prime\prime}$ = 0.081 $\pm$ 0.02 ($T$ = 6.7 K). 
There are relatively large differences among $\beta^{\prime\prime}$,
$b^{\prime}$, $b^{\prime\prime}$, and $a$, depending on the conditions. 
Nevertheless, we can say that these exponents are roughly the same and are
on the order of 0.05 - 0.08 around 7 K. The value of these exponents is
nearly equal to that of the exponent $y$ (=0.066 $\pm$ 0.001) derived from
the scaling analysis in Sec.~\ref{resultB}.  Note that our result of $a$ is
in good agreement with $a$ = 0.060 - 0.088 for the FM phase of the
reentrant ferromagnet
(Fe$_{0.65}$Ni$_{0.35}$)$_{0.882}$Mn$_{0.118}$.\cite{Roshko1994}

Finally we estimate the value of $y$.  The exponent $y$ is given by $y =
\beta/x$, where $x = \nu z$.  When we use the scaling relation, $\beta =
2\nu \theta$, which is derived in our previous paper,\cite{Suzuki2003} $y$
is given by $y = 2\theta/z$, where $\theta$ is the energy exponent and $z$
is the dynamic critical exponent.  The values of $\theta$ and $z$ are
theoretically predicted: $z$ = 6.0 $\pm$ 0.5 for the 3D $\pm J$ Ising spin
glass model (Ogielski\cite{Ogielski1985}) and $\theta$ = 0.19 $\pm$ 0.01
(Bray and Moore\cite{Bray1987}).  Using these values, $y$ is estimated as
$y$ = 0.063 $\pm$ 0.017, which is in good agreement with our result ($y$ =
0.066 $\pm$ 0.001).

\section{\label{conc}CONCLUSION}
The nonlinear susceptibility and aging phenomena are observed in the FM
phase of the reentrant ferromagnet Cu$_{0.2}$Co$_{0.8}$Cl$_{2}$-FeCl$_{3}$
GBIC. These results indicate that the FM phase between $T_{RSG}$ and
$T_{c}$ has a chaotic nature.  The absorption
$\chi^{\prime\prime}(t,\omega)$ is described by a power law form
$t^{-b^{\prime\prime}}$ but not by a $\omega t$-scaling form ($\omega
t)^{-b^{\prime\prime}}$.  A dynamic crossover behavior is observed around
$T_{RSG}$.  The time dependence of $M_{ZFC}$ has a stretched-exponential
form for the RSG phase, and a power-law form for the FM phase.  Further
studies on aging behaviors including memory effect, rejuvenation, and wait
time dependence, are required to understand the nature of the FM phase,
with the same qualitative features as in conventional spin glasses.

\begin{acknowledgments}
We would like to thank H. Suematsu for providing us with single crystal
kish graphite, T.-Y. Huang for his assistance in susceptibility
measurements, T. Shima and B. Olson for their assistance in sample
preparation and x-ray characterization.  Early work, in particular for the
sample preparation, was supported by NSF DMR 9201656.
\end{acknowledgments}

\end{document}